\documentclass[aps,prx,twocolumn,amsmath,amssymb,showpacs,superscriptaddress,notitlepage]{revtex4-1}
\usepackage{amsmath,amssymb,amsfonts,bm}
\usepackage{graphicx}
\usepackage{dcolumn}
\usepackage{mathrsfs}
\usepackage{bbold}
\usepackage{dsfont}
\usepackage{dcolumn}
\usepackage{epstopdf}
\usepackage[colorlinks=true,linkcolor=blue,citecolor=blue, urlcolor=blue,bookmarks=false]{hyperref}
\usepackage{changes}
\usepackage{textgreek}

\begin{document}

\title{Topological surface altermagnets in SSH-stacked magnetic layers}

\author{Rui Chen}
\affiliation{Department of Physics, Hubei University, Wuhan 430062, China}

\author{Bin Zhou}
\affiliation{Department of Physics, Hubei University, Wuhan 430062, China}

\author{Dong-Hui Xu}\email[]{donghuixu@cqu.edu.cn}
\affiliation{Department of Physics and Chongqing Key Laboratory for Strongly Coupled Physics, Chongqing University, Chongqing 400044, China}
\affiliation{Center of Quantum Materials and Devices, Chongqing University, Chongqing 400044, China}

\begin{abstract}
Surface altermagnetism opens new avenues in spintronics by unlocking altermagnetic spin-splitting at the boundaries of conventional antiferromagnets, bypassing the strict symmetry requirements of bulk altermagnets. In this work, we propose creating topological surface altermagnet by stacking magnetic layers in a Su-Schrieffer-Heeger pattern. We show that while the bulk of the system is a standard antiferromagnet with degenerate bands protected by $PT$ symmetry, breaking the local symmetry at the boundary gives rise to a topologically protected surface altermagnetic state residing within the topological gap. Furthermore, we propose that this effect can be experimentally detected by applying a perpendicular electric field. Besides, this approach can be readily generalized to surface altermagnetism of different types. Our work establishes topological boundaries as a natural platform for surface altermagnetism, offering a distinct route for realizing and manipulating topological surface altermagnets.
\end{abstract}
\maketitle

{\color{blue}\emph{Introduction}.}---As a newly identified category of magnetic materials, altermagnets differ significantly from traditional ferromagnetic and antiferromagnetic systems~\cite{Smejkal22PRX,Bai2024AFM,Naka19NC,Ahn19PRB,
hayami2019momentum,Yuan2020Giant,mazin2021prediction,
ma2021multifunctional,
ifmmode2022Emerging,krempasky2024altermagnetic,mejkal2022NatRevMat,LiS2026SCMA}. They exhibit a zero net magnetization and collinear spin arrangements typical of antiferromagnets, while simultaneously displaying the spin-polarized energy bands commonly associated with ferromagnets. Due to these unusual properties, altermagnets enable unique physical phenomena such as anomalous transport~\cite{Nakaprb,shao2021spin,Fernandestoplogical,zhang2024prl,Zhou2024PRL,ChenYY2025PRL}, giant tunneling magnetoresistance~\cite{sdongFeSb2,ifmmode2022Giant}, and potential topological responses~\cite{Ezawa2024PRLAlter,LiuCC2024PRB,ChenR2024altermagnet,ChenR2025altermagnet}, thereby prompting widespread research both theoretically and experimentally~\cite{Bai2023PRL,Gonzalez2021PRL,vsmejkal2020crystal,Feng2022NatElec, Fedchenko2024SciAdv,Karube2022PRL,Keler2024NpjS}.

\begin{figure}[ht!]
	\centering
	\includegraphics[width=\columnwidth]{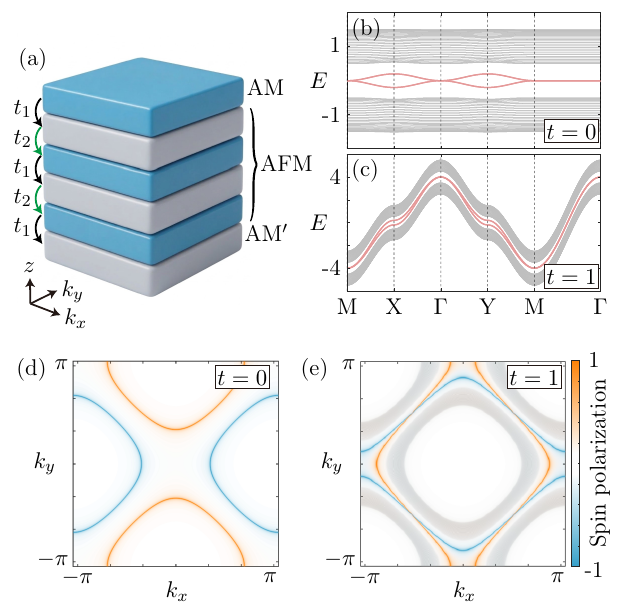}
	\caption{(a) Schematic illustration of the SSH stacking of two-dimensional magnetic layers. The alternating layers, colored by blue and gray, are coupled by alternating interlayer hoppings $t_1$ and $t_2$. The top surface displays the surface Brillouin zone with high-symmetry points $\Gamma$, $X$, $M$, and $Y$. (b), (c) Energy spectra of the finite layered system along high-symmetry paths for (b) vanishing intra-layer non-magnetic hopping $t = 0$  and (c) finite intra-layer hopping $t = 1$. Gray areas denote the bulk bands, while the red lines highlight the topologically protected surface states emerging within the bulk gap. (d), (e) The corresponding spin polarization of the surface local density of states (LDOS) in the $k_x$-$k_y$ momentum space for $t = 0$ (d) and $t = 1$ (e), where we take the Fermi energy $E_F=0.05$. }
	\label{fig_illustration}
		\vspace{0.8cm}
\end{figure}

Recently, the concept of surface altermagnetism has been introduced~\cite{Leeb2026arXiv,Lange2026arXiv,Sasioglu2026arXiv,HuYZ2026PRL}, characterized by systems where the bulk remains a conventional antiferromagnet while the boundary exhibits altermagnetism driven by local symmetry breaking.  The importance of surface altermagnetism lies in its capacity to bypass the stringent material constraints of bulk altermagnets. By utilizing the boundaries of widely available and well-established conventional antiferromagnets, it offers a highly accessible platform for low-dimensional altermagnets. Researchers have proposed surface altermagnetic states across various material platforms, including CuMnAs~\cite{Leeb2026arXiv}, NaMnP~\cite{Lange2026arXiv}, FeGe$_2$~\cite{Lange2026arXiv}, V$_3$Al~\cite{Sasioglu2026arXiv}, and BaMn$_2$Sb$_2$~\cite{Sasioglu2026arXiv}.


In this work, we propose an approach to realize topological surface altermagnet. By simply stacking two-dimensional magnetic layers in a Su-Schrieffer-Heeger (SSH) sequence~\cite{Su1979PRL}, we construct a structure where the bulk acts as a standard spin-degenerate antiferromagnet protected by $PT$ symmetry, where $P$ stands for inversion symmetry and $T$ stands for time-reversal symmetry. Crucially, the breaking of global symmetry at the open boundaries of this model intrinsically generates robust surface altermagnetic states within the topological gap.  Furthermore, we show that this emergent surface altermagnetism can be probed by applying a perpendicular electric field. Our findings thus provide not only a distinct topological mechanism for altermagnetism but also a versatile design principle for surface altermagnets.

{\color{blue}\emph{Model}.}--- We construct a minimal tight-binding model based on an SSH stacking~\cite{Su1979PRL} of magnetic layers [Fig.~\ref{fig_illustration}(a)], with
\begin{equation}
H(\mathbf{k})=\begin{bmatrix}h(\mathbf{k}_\parallel)&t_1+t_2e^{-ik_z}\\t_1+t_2e^{ik_z}&h^\prime(\mathbf{k}_\parallel)\end{bmatrix}.
\label{Eq:Model}
\end{equation}
Here, the diagonal blocks $h(\mathbf{k}_\parallel)$ and $h^\prime(\mathbf{k}_\parallel)$ describe the intralayer Hamiltonians of two adjacent magnetic layers. Crucially, the individual layer Hamiltonian $h(\mathbf{k}_\parallel)$ or $h^\prime(\mathbf{k}_\parallel)$ inherently breaks the $PT$ symmetry, hosting an altermagnetic state. The two layers act as $PT$ partners with opposite altermagnetic spin splittings. When coupled through the interlayer hoppings ($t_1$ and $t_2$), the combined unit cell regains the global $PT$ symmetry, rendering the bulk system a spin-degenerate antiferromagnet. Such a configuration is also referred to as a hidden altermagnet~\cite{GuoSD2026FronPhys,Matsuda2025PRL,Yang2025arXiv,LiChuang2026arXiv,HuYZ2026PRL1}, wherein the intrinsic altermagnetic signatures of individual layers are entirely masked by its $PT$ partner. This physical picture has been recently proposed in different systems, including Cr$_2$SO~\cite{GuoSD2026FronPhys} and Cs$_{1-\delta}$V$_2$Te$_2 $O~\cite{Yang2025arXiv}.
Experimentally, the alternating interlayer couplings $t_1$ and
$t_2$ can be engineered in artificial multilayers or van
der Waals heterostructures by intercalating non-magnetic
layers of varying thicknesses~\cite{Parkin1990PRL,Geim2013Nature,SongTC2018Science,Gibertini2019NatureNat}. 

For simplicity, we choose a standard two-dimensional square lattice model with a $d$-wave coupling, setting $h^{(\prime)}(\mathbf{k}_\parallel)=2t(\cos k_x+\cos k_y)s_0\pm2J(\cos k_x-\cos k_y)s_z$~\cite{Smejkal22PRX}. Lifting the $PT$ symmetry within such a framework provides a direct route to establishing a $d$-wave altermagnetic phase. Here, the parameter $J$ denotes the effective exchange coupling or the intrinsic magnetization within a single layer, and $t$ describes the intra-layer hopping strength. In the following calculations, we fix $t_1=0.5$, $t_2=1$, $J=0.05$.



{\color{blue}\emph{The emergent surface altermagnetism}.}---Truncating this system at the open boundaries inherently disrupts the local $PT$ symmetry. This symmetry breaking can be intuitively understood in the fully dimerized limit of $t_1 = 0$, as illustrated in Fig.~\ref{fig_illustration}(a). In this extreme regime, the bulk simply forms isolated two-layer 
dimers coupled solely by $t_2$. Within each bulk dimer, the pairing 
of two adjacent altermagnetic layers perfectly preserves the local 
$PT$ symmetry. However, this periodic dimerization  leaves a completely 
decoupled, uncompensated single altermagnetic layer at the surface. This isolated surface 
layer breaks the local $PT$ symmetry, thereby 
constituting a surface altermagnet.

Remarkably, owing to topological protection, this surface altermagnet remains robust even away from the fully dimerized limit when $t_1 \neq 0$~\cite{Su1979PRL,Shen2017TI,Asbth2016Note}.  Figure~\ref{fig_illustration}(b) presents the energy spectrum of the system for a finite $t_1=0.5$, where we consider the simplest case with vanishing intra-layer hopping ($t=0$). We observe that the system opens a finite bulk gap, and a set of distinct boundary states emerges within this gap.  By checking the spin-resolved surface local density of states (LDOS) in Fig.~\ref{fig_illustration}(d), we find that the  surface states are isolated from the bulk bands and exhibit strong spin separation, reflecting an altermagnetic behavior.

This scenario is explained as follows. Because the in-plane momentum $\mathbf{k}_\parallel$ is a conserved good quantum number, the full Hamiltonian in Eq.~\eqref{Eq:Model} can be decoupled into a continuous family of independent SSH models defined at each $\mathbf{k}_\parallel$. Specifically, along the M-$\Gamma$ line in Fig.~\ref{fig_illustration}(b), where $h^{(\prime)}(\mathbf{k}_\parallel)=0$, the system reduces to the pristine SSH model, supporting exact zero-energy topological boundary modes. At other $\mathbf{k}_\parallel$ points, the term $\pm 2J(\cos k_x - \cos k_y)s_z$ breaks the chiral symmetry of the pristine SSH model, shifting these boundary modes away from zero energy by an exact amplitude of $\pm 2J(\cos k_x - \cos k_y)$ (see Sec. SI of Supplementary Material~\cite{Supp}). Consequently, the continuous collection of these shifted boundary modes across all $\mathbf{k}_\parallel$ points collectively forms a robust $d$-wave surface altermagnet localized within the bulk gap.


This physical picture naturally extends to the general regime with
non-vanishing intra-layer hopping ($t \neq 0$) 
[Fig.~\ref{fig_illustration}(c)], where the surface 
altermagnetism survives within the gap. The $t$ term manifests in the Hamiltonian simply as $2t(\cos k_x + \cos k_y)s_0$. This term acts merely as a momentum-dependent, uniform energy shift for the independent SSH chain at any given $\mathbf{k}_\parallel$ point. Therefore, while this energy shift introduces an overall band dispersion that may eventually close the global band gap of the system, it strictly preserves the direct band gap at each specific momentum point. Consequently, the emergent surface altermagnetic states remain perfectly well-defined and are robustly protected within this direct bulk gap.  This scenario can be observed more clearly in Fig.~\ref{fig_illustration}(e), the strongly spin-resolved altermagnetic surface states are clearly separated from the degenerate bulk-projected states, which act as an unpolarized background.

{\color{blue}\emph{Topological nature of the surface altermagnet}.}---The topological character of the system is most transparently defined along the specific high-symmetry lines $k_x = \pm k_y$, where the local magnetic order effectively vanishes. At these nodes, the topology is characterized by the Zak phase associated with chiral symmetry, strictly quantized to either $0$ or $\pi$ depending on the interlayer coupling ratio $t_1/t_2$. This quantization characterizes the topologically protected  boundary modes~\cite{Bernevig2013TI,Hasan2010RMP,Qi2011RMP}. Moving away from these specific points to generic momenta in the two-dimensional Brillouin zone, the boundary states shift to finite energies. However,  these finite-energy states are adiabatically connected to the topologically protected zero-energy modes, continuously stitching together across the Brillouin zone to form a robust surface altermagnet.

In Sec.~SIII of the Supplementary Material~\cite{Supp}, we calculate the energy spectrum for the topologically trivial regime with $t_1 > t_2$, where the system corresponds to a trivial phase.  As expected, the surface 
altermagnetic states completely disappear in this case, further 
demonstrating that the emergence of the surface altermagnetism is 
fundamentally rooted in the non-trivial bulk topology rather than 
being a trivial boundary artifact. 

\begin{figure}[t!]
	\centering
	\includegraphics[width=\columnwidth]{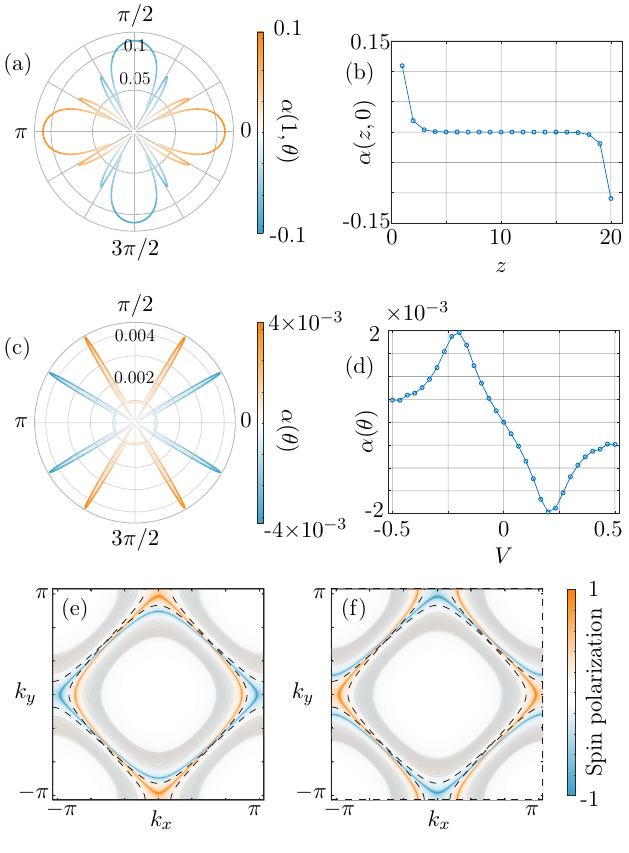}
	\caption{Transport signatures and surface spin polarization of the system 
		in the absence (a)-(b) and presence (e)-(f) of a perpendicular 
		electric field. (a) Angle-dependent spin splitter angle $\alpha(1, \theta)$ 
		for the outermost surface layer. (b) Layer-resolved spin splitter 
		angle $\alpha(z, 0)$ across the finite slab. 
		(c) Angle-dependent total spin splitter angle $\alpha(\theta)$, (d) evolution of the total spin splitter angle 
		$\alpha(\theta)$ with respect to the strength of the external electric field $V$, and (e)-(f) 
		spin-resolved surface local density of states in the momentum 
		space. In (e) and (f), the dashed lines denote the contours of the 
		spin-polarized surface states at $V=0$, extracted from 
		Fig.~\ref{fig_illustration}(e).}
	\label{fig_conductance}
	\vspace{-0.3 cm}
\end{figure}

{\color{blue}\emph{Layer-resolved spin splitter effect}.}---
To characterize the transport signatures of the emergent
surface altermagnetism, we calculate the layer-resolved spin splitter
angle~\cite{Leeb2026arXiv,Gonzalez2021PRL}, which quantifies the degree
of spin splitting by measuring the ratio of the spin conductivity to
the charge conductivity under an applied in-plane electric field.
The layer-dependent spin splitter angle is defined as $\alpha(z,\theta) = 2 \arctan \left( \frac{\sigma_{xx}^\uparrow(z,\theta) - \sigma_{xx}^\downarrow(z,\theta)}{\sigma_{xx}^\uparrow(z,\theta) + \sigma_{xx}^\downarrow(z,\theta)} \right)$,
where $\sigma_{xx}^{\uparrow}(z,\theta)$ and $\sigma_{xx}^{\downarrow}(z,\theta)$ represent 
the spin-resolved longitudinal conductivities for $j$-th layer along the angle $\theta$ [see Sec. SII of Supplementary Material~\cite{Supp} for more details].  Furthermore, to evaluate the macroscopic spin transport behavior of 
the entire system, we calculate the total spin splitter angle 
$\alpha(\theta)$ by summing over all layers, defined as: $\alpha(\theta) = \sum_z \alpha(z,\theta)$.

As illustrated in Fig.~\ref{fig_conductance}(a), we calculate 
spin splitter angle for the surface layer, denoted as $\alpha(1,\theta)$. 
A pronounced, highly anisotropic $d$-wave spin polarization signal is 
clearly observed at the surface, reflecting the intrinsic 
altermagnetic symmetry of the uncompensated boundary layer. Figure~\ref{fig_conductance}(b) displays the 
layer-resolved spin splitter angle $\alpha(z,0)$ across different layers $z$ with $\theta=0$. The calculated spin polarization is found to be highly 
localized at the two opposite boundaries, rapidly decaying to 
zero within the bulk region.  Notably, the spin splitter angles at the top and bottom surfaces exhibit opposite signs but equal magnitudes. This spatial profile confirms that the observed altermagnetic response stems from the uncompensated boundary layers that are connected by $PT$ symmetry. The bulk interior remains perfectly compensated, yielding 
a vanishing spin polarization due to the preservation of the local $PT$ symmetry between adjacent layers.

{\color{blue}\emph{Experimental detection via an external perpendicular electric field}.}---
While the layer-resolved spin splitter angle provides a clear theoretical
signature, isolating the precise transport contribution from a single
atomic layer remains highly challenging in practical experiments. To overcome this limitation and facilitate direct experimental 
detection, we propose applying a perpendicular electric field 
of strength $V$, which introduces a layer-dependent linear 
potential $V_z = V \frac{2z - N_z - 1}{N_z - 1}$ for the $z$-th 
layer in a finite slab of $N_z$ layers.

As shown in Fig.~\ref{fig_conductance}(d), the total 
spin splitter angle $\alpha$ is strictly zero at $V=0$ due to the 
perfect compensation between the opposite surfaces. Upon applying 
a finite $V$, a net macroscopic spin polarization emerges, and 
its magnitude exhibits an odd-function dependence on the electric 
field strength. Figure~\ref{fig_conductance}(c) presents the 
angle-dependent total spin splitter angle at a finite $V$, which 
displays a strongly anisotropic response.

The microscopic origin of this field-induced net polarization 
is further elucidated by the spin-resolved surface local 
density of states in Fig.~\ref{fig_conductance}(e) and (f). 
Compared to the unperturbed surface states at $V=0$ (indicated 
by the dashed lines), the perpendicular electric field breaks $PT$ symmetry and introduces 
an opposite electrostatic potential at the two boundaries. This 
perturbation expands or shrinks the Fermi contours of the 
spin-polarized surface states in the momentum space. Consequently, 
the exact cancellation of the spin responses from the top and 
bottom surfaces is destroyed, yielding the non-zero macroscopic 
transport signatures. 

Moreover, the odd 
behavior in Fig.~\ref{fig_conductance}(d) originates from the opposite spin-polarization nature of the 
top and bottom surface layers. Reversing the direction of the electric 
field ($V \rightarrow -V$) exactly exchanges the electrostatic potentials 
applied to these two boundaries. Since the opposite surfaces inherently 
host altermagnetic states with opposite spins, reversing the potential 
gradient naturally flips the dominant spin carrier of the system, 
thereby reversing the sign of the net macroscopic spin response.

It is noteworthy
that an analogous  approach has already been
employed to detect the layer Hall effect in antiferromagnetic topological
insulators~\cite{Gao2021Nature,ChenR2022NSR}, thereby establishing the experimental viability of
this technique for resolving layer-locked transport phenomena. Furthermore, recent experiments
have achieved remarkable sensitivities in measuring spin splitter angles,
capable of resolving signals about $10^{-3}$~\cite{Jechumtal2026PRB}. Consequently, the
macroscopic spin splitting signal predicted in our field-driven
architecture falls well within the resolution limits of state-of-the-art
transport measurements. 

\begin{figure}[t!]
	\centering
	\includegraphics[width=\columnwidth]{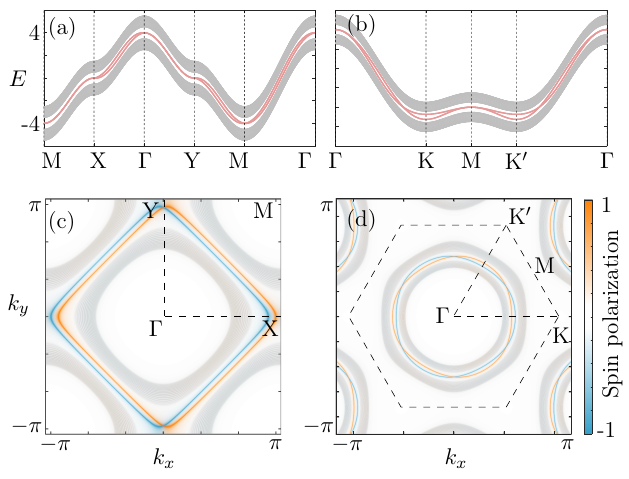}
	\caption{Generalization of topological surface altermagnetism to $p$-wave and $f$-wave symmetries. (a, b) Projected one-dimensional energy band structures for the $p$-wave (a) and $f$-wave (b) altermagnetic systems. The projected bulk bands are depicted in gray, while the topologically protected surface states emerging within the direct bulk gap are highlighted in red. (c, d) The corresponding spin-polarized surface local density of states in momentum space for the $p$-wave (c) and $f$-wave (d) configurations. In (d), the dashed lines outline the Brillouin zone boundaries. }
	\label{fig_pfwave}
	\vspace{-0.3 cm}
\end{figure}

{\color{blue}\emph{Generalization to more scenarios}.}---
The physical mechanism of topological surface altermagnetism proposed
herein is highly versatile and not restricted to the specific $d$-wave
configuration discussed above. To demonstrate the universality of our
design principle, we generalize the SSH-like stacking architecture
to encompass 2D magnetic layers exhibiting other unconventional spin
textures, such as $p$-wave and $g$-wave altermagnetism. By substituting the intra-layer Hamiltonian with the corresponding 
$p$-wave model $h_p(\mathbf{k}_\parallel) = 2t(\cos k_x+\cos k_y)s_0 + 2J\sin k_x s_z$ 
or the $f$-wave model $h_f(\mathbf{k}_\parallel) = \left[ 2t\cos k_x + 4t\cos(k_x/2)\cos(\sqrt{3}k_y/2) \right]s_0 + 2J\sin k_x (\cos k_x - \cos(\sqrt{3}k_y))s_z$, 
we investigate the emergent surface altermagnetism in the $p$- and $f$-wave systems~\cite{Smejkal22PRX,Hayami2020PRB}.

As shown in Fig.~\ref{fig_pfwave}, we calculate the spectra and the spin-resolved surface
LDOS for the two systems, and observe similar results. In each case, the truncation of the periodic chain safely
preserves the local in-plane spin-momentum locking while simultaneously
breaking the global $PT$ symmetry at the open boundaries.
Consequently, strongly spin-resolved, uncompensated surface states
emerge perfectly localized within the topological bulk gap. This
confirms that the emergence of surface altermagnetism driven by
geometrical boundary truncation is a robust and universally applicable
phenomenon, offering a broad material and symmetry platform for
designing topological surface altermagnetism.

{\color{blue}\emph{Conclusion and Discussion}.}---
In summary, we have proposed a paradigm for
realizing and manipulating topological surface altermagnetism. By
stacking two-dimensional magnetic layers in an 
SSH sequence, we demonstrated that the geometrical
truncation at the open boundaries intrinsically breaks the global
$PT$ symmetry of the bulk antiferromagnet. This symmetry
reduction naturally hosts robust, uncompensated altermagnetic states
localized at the surface, which are topologically protected within
the direct bulk gap. Furthermore, we show that applying a
perpendicular electric field provides an approach to
detect these boundary states via a macroscopic, tunable spin
splitter angle, and we systematically generalized this mechanism
across different types of altermagnetism.


\begin{acknowledgments}
	This work was supported by the NSFC (under Grants Nos.~12474151, 92565103, 12304195, 12547101, and U25D8012). D.H.X. acknowledges the Beijing National Laboratory for Condensed Matter Physics (No. 2024BNLCMPKF025) and the Fundamental Research Funds for the Central Universities (Grant No. 2025CDJIAISYB-032). R.C. acknowledges the Chutian Scholars Program in Hubei Province, the Hubei Provincial Natural Science Foundation (Grant No. 2025AFA081), the Guangdong Provincial Quantum Science Strategic Initiative (GDZX2401001), the key project of Hubei provincial department of education (under Grant No. D20241004), and the original seed program of Hubei University. B.Z. acknowledges the Wuhan city key R\&D program (under Grant No. 2025050602030069).
\end{acknowledgments}

%
%
%
\bibliographystyle{apsrev4-1-etal-title_6authors}
\bibliography{refs-transport,refs-transport_v1}

\begin{thebibliography}{55}%
\makeatletter
\providecommand \@ifxundefined [1]{%
 \@ifx{#1\undefined}
}%
\providecommand \@ifnum [1]{%
 \ifnum #1\expandafter \@firstoftwo
 \else \expandafter \@secondoftwo
 \fi
}%
\providecommand \@ifx [1]{%
 \ifx #1\expandafter \@firstoftwo
 \else \expandafter \@secondoftwo
 \fi
}%
\providecommand \natexlab [1]{#1}%
\providecommand \enquote  [1]{``#1''}%
\providecommand \bibnamefont  [1]{#1}%
\providecommand \bibfnamefont [1]{#1}%
\providecommand \citenamefont [1]{#1}%
\providecommand \href@noop [0]{\@secondoftwo}%
\providecommand \href [0]{\begingroup \@sanitize@url \@href}%
\providecommand \@href[1]{\@@startlink{#1}\@@href}%
\providecommand \@@href[1]{\endgroup#1\@@endlink}%
\providecommand \@sanitize@url [0]{\catcode `\\12\catcode `\$12\catcode
  `\&12\catcode `\#12\catcode `\^12\catcode `\_12\catcode `\%12\relax}%
\providecommand \@@startlink[1]{}%
\providecommand \@@endlink[0]{}%
\providecommand \url  [0]{\begingroup\@sanitize@url \@url }%
\providecommand \@url [1]{\endgroup\@href {#1}{\urlprefix }}%
\providecommand \urlprefix  [0]{URL }%
\providecommand \Eprint [0]{\href }%
\providecommand \doibase [0]{http://dx.doi.org/}%
\providecommand \selectlanguage [0]{\@gobble}%
\providecommand \bibinfo  [0]{\@secondoftwo}%
\providecommand \bibfield  [0]{\@secondoftwo}%
\providecommand \translation [1]{[#1]}%
\providecommand \BibitemOpen [0]{}%
\providecommand \bibitemStop [0]{}%
\providecommand \bibitemNoStop [0]{.\EOS\space}%
\providecommand \EOS [0]{\spacefactor3000\relax}%
\providecommand \BibitemShut  [1]{\csname bibitem#1\endcsname}%
\let\auto@bib@innerbib\@empty
\bibitem [{\citenamefont {\ifmmode~\check{S}\else \v{S}\fi{}mejkal}\ \emph
  {et~al.}(2022{\natexlab{a}})\citenamefont {\ifmmode~\check{S}\else
  \v{S}\fi{}mejkal}, \citenamefont {Sinova},\ and\ \citenamefont
  {Jungwirth}}]{Smejkal22PRX}%
  \BibitemOpen
  \bibfield  {author} {\bibinfo {author} {\bibfnamefont {L.}~\bibnamefont
  {\ifmmode~\check{S}\else \v{S}\fi{}mejkal}}, \bibinfo {author} {\bibfnamefont
  {J.}~\bibnamefont {Sinova}}, \ and\ \bibinfo {author} {\bibfnamefont
  {T.}~\bibnamefont {Jungwirth}},\ }\bibfield  {title} {\enquote {\bibinfo
  {title} {Beyond conventional ferromagnetism and antiferromagnetism: A phase
  with nonrelativistic spin and crystal rotation symmetry}}, }\href {\doibase
  10.1103/PhysRevX.12.031042} {\bibfield  {journal} {\bibinfo  {journal} {Phys.
  Rev. X}\ }\textbf {\bibinfo {volume} {12}},\ \bibinfo {pages} {031042}
  (\bibinfo {year} {2022}{\natexlab{a}})}\BibitemShut {NoStop}%
\bibitem [{\citenamefont {Bai}\ \emph {et~al.}(2024)\citenamefont {Bai},
  \citenamefont {Feng}, \citenamefont {Liu}, \citenamefont {Šmejkal},
  \citenamefont {Mokrousov},\ and\ \citenamefont {Yao}}]{Bai2024AFM}%
  \BibitemOpen
  \bibfield  {author} {\bibinfo {author} {\bibfnamefont {L.}~\bibnamefont
  {Bai}}, \bibinfo {author} {\bibfnamefont {W.}~\bibnamefont {Feng}}, \bibinfo
  {author} {\bibfnamefont {S.}~\bibnamefont {Liu}}, \bibinfo {author}
  {\bibfnamefont {L.}~\bibnamefont {Šmejkal}}, \bibinfo {author}
  {\bibfnamefont {Y.}~\bibnamefont {Mokrousov}}, \ and\ \bibinfo {author}
  {\bibfnamefont {Y.}~\bibnamefont {Yao}},\ }\bibfield  {title} {\enquote
  {\bibinfo {title} {Altermagnetism: Exploring new frontiers in magnetism and
  spintronics}}, }\href {\doibase 10.1002/adfm.202409327} {\bibfield  {journal}
  {\bibinfo  {journal} {Adv. Funct. Mater.}\ ,\ \bibinfo {pages} {2409327}}
  (\bibinfo {year} {2024})}\BibitemShut {NoStop}%
\bibitem [{\citenamefont {Naka}\ \emph {et~al.}(2019)\citenamefont {Naka},
  \citenamefont {Hayami}, \citenamefont {Kusunose}, \citenamefont {Yanagi},
  \citenamefont {Motome},\ and\ \citenamefont {Seo}}]{Naka19NC}%
  \BibitemOpen
  \bibfield  {author} {\bibinfo {author} {\bibfnamefont {M.}~\bibnamefont
  {Naka}}, \bibinfo {author} {\bibfnamefont {S.}~\bibnamefont {Hayami}},
  \bibinfo {author} {\bibfnamefont {H.}~\bibnamefont {Kusunose}}, \bibinfo
  {author} {\bibfnamefont {Y.}~\bibnamefont {Yanagi}}, \bibinfo {author}
  {\bibfnamefont {Y.}~\bibnamefont {Motome}}, \ and\ \bibinfo {author}
  {\bibfnamefont {H.}~\bibnamefont {Seo}},\ }\bibfield  {title} {\enquote
  {\bibinfo {title} {Spin current generation in organic antiferromagnets}},
  }\href {\doibase 10.1038/s41467-019-12229-y} {\bibfield  {journal} {\bibinfo
  {journal} {Nat. Commun.}\ }\textbf {\bibinfo {volume} {10}},\ \bibinfo
  {pages} {4305} (\bibinfo {year} {2019})}\BibitemShut {NoStop}%
\bibitem [{\citenamefont {Ahn}\ \emph {et~al.}(2019)\citenamefont {Ahn},
  \citenamefont {Hariki}, \citenamefont {Lee},\ and\ \citenamefont
  {Kune\ifmmode~\check{s}\else \v{s}\fi{}}}]{Ahn19PRB}%
  \BibitemOpen
  \bibfield  {author} {\bibinfo {author} {\bibfnamefont {K.-H.}\ \bibnamefont
  {Ahn}}, \bibinfo {author} {\bibfnamefont {A.}~\bibnamefont {Hariki}},
  \bibinfo {author} {\bibfnamefont {K.-W.}\ \bibnamefont {Lee}}, \ and\
  \bibinfo {author} {\bibfnamefont {J.}~\bibnamefont
  {Kune\ifmmode~\check{s}\else \v{s}\fi{}}},\ }\bibfield  {title} {\enquote
  {\bibinfo {title} {{Antiferromagnetism in ${\mathrm{RuO}}_{2}$ as $d$-wave
  Pomeranchuk instability}}}, }\href {\doibase 10.1103/PhysRevB.99.184432}
  {\bibfield  {journal} {\bibinfo  {journal} {Phys. Rev. B}\ }\textbf {\bibinfo
  {volume} {99}},\ \bibinfo {pages} {184432} (\bibinfo {year}
  {2019})}\BibitemShut {NoStop}%
\bibitem [{\citenamefont {Hayami}\ \emph {et~al.}(2019)\citenamefont {Hayami},
  \citenamefont {Yanagi},\ and\ \citenamefont {Kusunose}}]{hayami2019momentum}%
  \BibitemOpen
  \bibfield  {author} {\bibinfo {author} {\bibfnamefont {S.}~\bibnamefont
  {Hayami}}, \bibinfo {author} {\bibfnamefont {Y.}~\bibnamefont {Yanagi}}, \
  and\ \bibinfo {author} {\bibfnamefont {H.}~\bibnamefont {Kusunose}},\
  }\bibfield  {title} {\enquote {\bibinfo {title} {Momentum-dependent spin
  splitting by collinear antiferromagnetic ordering}}, }\href
  {https://doi.org/10.7566/JPSJ.88.123702} {\bibfield  {journal} {\bibinfo
  {journal} {J. Phys. Soc. Jpn.}\ }\textbf {\bibinfo {volume} {88}},\ \bibinfo
  {pages} {123702} (\bibinfo {year} {2019})}\BibitemShut {NoStop}%
\bibitem [{\citenamefont {Yuan}\ \emph {et~al.}(2020)\citenamefont {Yuan},
  \citenamefont {Wang}, \citenamefont {Luo}, \citenamefont {Rashba},\ and\
  \citenamefont {Zunger}}]{Yuan2020Giant}%
  \BibitemOpen
  \bibfield  {author} {\bibinfo {author} {\bibfnamefont {L.-D.}\ \bibnamefont
  {Yuan}}, \bibinfo {author} {\bibfnamefont {Z.}~\bibnamefont {Wang}}, \bibinfo
  {author} {\bibfnamefont {J.-W.}\ \bibnamefont {Luo}}, \bibinfo {author}
  {\bibfnamefont {E.~I.}\ \bibnamefont {Rashba}}, \ and\ \bibinfo {author}
  {\bibfnamefont {A.}~\bibnamefont {Zunger}},\ }\bibfield  {title} {\enquote
  {\bibinfo {title} {Giant momentum-dependent spin splitting in centrosymmetric
  low-$z$ antiferromagnets}}, }\href {\doibase 10.1103/PhysRevB.102.014422}
  {\bibfield  {journal} {\bibinfo  {journal} {Phys. Rev. B}\ }\textbf {\bibinfo
  {volume} {102}},\ \bibinfo {pages} {014422} (\bibinfo {year}
  {2020})}\BibitemShut {NoStop}%
\bibitem [{\citenamefont {Mazin}\ \emph {et~al.}(2021)\citenamefont {Mazin},
  \citenamefont {Koepernik}, \citenamefont {Johannes}, \citenamefont
  {Gonz{\'a}lez-Hern{\'a}ndez},\ and\ \citenamefont
  {{\v{S}}mejkal}}]{mazin2021prediction}%
  \BibitemOpen
  \bibfield  {author} {\bibinfo {author} {\bibfnamefont {I.~I.}\ \bibnamefont
  {Mazin}}, \bibinfo {author} {\bibfnamefont {K.}~\bibnamefont {Koepernik}},
  \bibinfo {author} {\bibfnamefont {M.~D.}\ \bibnamefont {Johannes}}, \bibinfo
  {author} {\bibfnamefont {R.}~\bibnamefont {Gonz{\'a}lez-Hern{\'a}ndez}}, \
  and\ \bibinfo {author} {\bibfnamefont {L.}~\bibnamefont {{\v{S}}mejkal}},\
  }\bibfield  {title} {\enquote {\bibinfo {title} {{Prediction of
  unconventional magnetism in doped FeSb$_2$}}}, }\href
  {http://dx.doi.org/10.1073/pnas.2108924118} {\bibfield  {journal} {\bibinfo
  {journal} {Proc. Nat. Acad. Sci.}\ }\textbf {\bibinfo {volume} {118}},\
  \bibinfo {pages} {e2108924118} (\bibinfo {year} {2021})}\BibitemShut
  {NoStop}%
\bibitem [{\citenamefont {Ma}\ \emph {et~al.}(2021)\citenamefont {Ma},
  \citenamefont {Hu}, \citenamefont {Li}, \citenamefont {Liu}, \citenamefont
  {Yao}, \citenamefont {Jia},\ and\ \citenamefont
  {Liu}}]{ma2021multifunctional}%
  \BibitemOpen
  \bibfield  {author} {\bibinfo {author} {\bibfnamefont {H.-Y.}\ \bibnamefont
  {Ma}}, \bibinfo {author} {\bibfnamefont {M.}~\bibnamefont {Hu}}, \bibinfo
  {author} {\bibfnamefont {N.}~\bibnamefont {Li}}, \bibinfo {author}
  {\bibfnamefont {J.}~\bibnamefont {Liu}}, \bibinfo {author} {\bibfnamefont
  {W.}~\bibnamefont {Yao}}, \bibinfo {author} {\bibfnamefont {J.-F.}\
  \bibnamefont {Jia}}, \ and\ \bibinfo {author} {\bibfnamefont
  {J.}~\bibnamefont {Liu}},\ }\bibfield  {title} {\enquote {\bibinfo {title}
  {Multifunctional antiferromagnetic materials with giant piezomagnetism and
  noncollinear spin current}}, }\href
  {https://doi.org/10.1038/s41467-021-23127-7} {\bibfield  {journal} {\bibinfo
  {journal} {Nat. Commun.}\ }\textbf {\bibinfo {volume} {12}},\ \bibinfo
  {pages} {2846} (\bibinfo {year} {2021})}\BibitemShut {NoStop}%
\bibitem [{\citenamefont {\ifmmode~\check{S}\else \v{S}\fi{}mejkal}\ \emph
  {et~al.}(2022{\natexlab{b}})\citenamefont {\ifmmode~\check{S}\else
  \v{S}\fi{}mejkal}, \citenamefont {Sinova},\ and\ \citenamefont
  {Jungwirth}}]{ifmmode2022Emerging}%
  \BibitemOpen
  \bibfield  {author} {\bibinfo {author} {\bibfnamefont {L.}~\bibnamefont
  {\ifmmode~\check{S}\else \v{S}\fi{}mejkal}}, \bibinfo {author} {\bibfnamefont
  {J.}~\bibnamefont {Sinova}}, \ and\ \bibinfo {author} {\bibfnamefont
  {T.}~\bibnamefont {Jungwirth}},\ }\bibfield  {title} {\enquote {\bibinfo
  {title} {Emerging research landscape of altermagnetism}}, }\href {\doibase
  10.1103/PhysRevX.12.040501} {\bibfield  {journal} {\bibinfo  {journal} {Phys.
  Rev. X}\ }\textbf {\bibinfo {volume} {12}},\ \bibinfo {pages} {040501}
  (\bibinfo {year} {2022}{\natexlab{b}})}\BibitemShut {NoStop}%
\bibitem [{\citenamefont {Krempask{\`y}}\ \emph {et~al.}(2024)\citenamefont
  {Krempask{\`y}}, \citenamefont {{\v{S}}mejkal}, \citenamefont {D’souza},
  \citenamefont {Hajlaoui}, \citenamefont {Springholz}, \citenamefont
  {Uhl{\'\i}{\v{r}}ov{\'a}}, \citenamefont {Alarab}, \citenamefont
  {Constantinou}, \citenamefont {Strocov}, \citenamefont {Usanov} \emph
  {et~al.}}]{krempasky2024altermagnetic}%
  \BibitemOpen
  \bibfield  {author} {\bibinfo {author} {\bibfnamefont {J.}~\bibnamefont
  {Krempask{\`y}}}, \bibinfo {author} {\bibfnamefont {L.}~\bibnamefont
  {{\v{S}}mejkal}}, \bibinfo {author} {\bibfnamefont {S.}~\bibnamefont
  {D’souza}}, \bibinfo {author} {\bibfnamefont {M.}~\bibnamefont {Hajlaoui}},
  \bibinfo {author} {\bibfnamefont {G.}~\bibnamefont {Springholz}}, \bibinfo
  {author} {\bibfnamefont {K.}~\bibnamefont {Uhl{\'\i}{\v{r}}ov{\'a}}},  \emph
  {et~al.},\ }\bibfield  {title} {\enquote {\bibinfo {title} {Altermagnetic
  lifting of kramers spin degeneracy}}, }\href
  {https://doi.org/10.1038/s41586-023-06907-7} {\bibfield  {journal} {\bibinfo
  {journal} {Nature}\ }\textbf {\bibinfo {volume} {626}},\ \bibinfo {pages}
  {517} (\bibinfo {year} {2024})}\BibitemShut {NoStop}%
\bibitem [{\citenamefont {Smejkal}\ \emph {et~al.}(2022)\citenamefont
  {Smejkal}, \citenamefont {MacDonald}, \citenamefont {Sinova}, \citenamefont
  {Nakatsuji},\ and\ \citenamefont {Jungwirth}}]{mejkal2022NatRevMat}%
  \BibitemOpen
  \bibfield  {author} {\bibinfo {author} {\bibfnamefont {L.}~\bibnamefont
  {Smejkal}}, \bibinfo {author} {\bibfnamefont {A.~H.}\ \bibnamefont
  {MacDonald}}, \bibinfo {author} {\bibfnamefont {J.}~\bibnamefont {Sinova}},
  \bibinfo {author} {\bibfnamefont {S.}~\bibnamefont {Nakatsuji}}, \ and\
  \bibinfo {author} {\bibfnamefont {T.}~\bibnamefont {Jungwirth}},\ }\bibfield
  {title} {\enquote {\bibinfo {title} {{Anomalous Hall antiferromagnets}}},
  }\href {\doibase 10.1038/s41578-022-00430-3} {\bibfield  {journal} {\bibinfo
  {journal} {Nat. Rev. Mat.}\ }\textbf {\bibinfo {volume} {7}},\ \bibinfo
  {pages} {482} (\bibinfo {year} {2022})}\BibitemShut {NoStop}%
\bibitem [{\citenamefont {Li}\ \emph {et~al.}(2026{\natexlab{a}})\citenamefont
  {Li}, \citenamefont {Chen}, \citenamefont {Pan}, \citenamefont {Li},
  \citenamefont {Zhang},\ and\ \citenamefont {Lu}}]{LiS2026SCMA}%
  \BibitemOpen
  \bibfield  {author} {\bibinfo {author} {\bibfnamefont {Y.-X.}\ \bibnamefont
  {Li}}, \bibinfo {author} {\bibfnamefont {Y.}~\bibnamefont {Chen}}, \bibinfo
  {author} {\bibfnamefont {L.}~\bibnamefont {Pan}}, \bibinfo {author}
  {\bibfnamefont {S.}~\bibnamefont {Li}}, \bibinfo {author} {\bibfnamefont
  {S.-B.}\ \bibnamefont {Zhang}}, \ and\ \bibinfo {author} {\bibfnamefont
  {H.-Z.}\ \bibnamefont {Lu}},\ }\bibfield  {title} {\enquote {\bibinfo {title}
  {Exploration of altermagnetism in {RuO$_2$}}}, }\href {\doibase
  10.1007/s11433-026-2913-8} {\bibfield  {journal} {\bibinfo  {journal} {Sci.
  China Phys. Mech. Astron.}\ }\textbf {\bibinfo {volume} {69}},\ \bibinfo
  {pages} {257001} (\bibinfo {year} {2026}{\natexlab{a}})}\BibitemShut
  {NoStop}%
\bibitem [{\citenamefont {Naka}\ \emph {et~al.}(2021)\citenamefont {Naka},
  \citenamefont {Motome},\ and\ \citenamefont {Seo}}]{Nakaprb}%
  \BibitemOpen
  \bibfield  {author} {\bibinfo {author} {\bibfnamefont {M.}~\bibnamefont
  {Naka}}, \bibinfo {author} {\bibfnamefont {Y.}~\bibnamefont {Motome}}, \ and\
  \bibinfo {author} {\bibfnamefont {H.}~\bibnamefont {Seo}},\ }\bibfield
  {title} {\enquote {\bibinfo {title} {Perovskite as a spin current
  generator}}, }\href {\doibase 10.1103/PhysRevB.103.125114} {\bibfield
  {journal} {\bibinfo  {journal} {Phys. Rev. B}\ }\textbf {\bibinfo {volume}
  {103}},\ \bibinfo {pages} {125114} (\bibinfo {year} {2021})}\BibitemShut
  {NoStop}%
\bibitem [{\citenamefont {Shao}\ \emph {et~al.}(2021)\citenamefont {Shao},
  \citenamefont {Zhang}, \citenamefont {Li}, \citenamefont {Eom},\ and\
  \citenamefont {Tsymbal}}]{shao2021spin}%
  \BibitemOpen
  \bibfield  {author} {\bibinfo {author} {\bibfnamefont {D.-F.}\ \bibnamefont
  {Shao}}, \bibinfo {author} {\bibfnamefont {S.-H.}\ \bibnamefont {Zhang}},
  \bibinfo {author} {\bibfnamefont {M.}~\bibnamefont {Li}}, \bibinfo {author}
  {\bibfnamefont {C.-B.}\ \bibnamefont {Eom}}, \ and\ \bibinfo {author}
  {\bibfnamefont {E.~Y.}\ \bibnamefont {Tsymbal}},\ }\bibfield  {title}
  {\enquote {\bibinfo {title} {Spin-neutral currents for spintronics}}, }\href
  {https://doi.org/10.1038/s41467-021-26915-3} {\bibfield  {journal} {\bibinfo
  {journal} {Nat. Commun.}\ }\textbf {\bibinfo {volume} {12}},\ \bibinfo
  {pages} {7061} (\bibinfo {year} {2021})}\BibitemShut {NoStop}%
\bibitem [{\citenamefont {Fernandes}\ \emph {et~al.}(2024)\citenamefont
  {Fernandes}, \citenamefont {de~Carvalho}, \citenamefont {Birol},\ and\
  \citenamefont {Pereira}}]{Fernandestoplogical}%
  \BibitemOpen
  \bibfield  {author} {\bibinfo {author} {\bibfnamefont {R.~M.}\ \bibnamefont
  {Fernandes}}, \bibinfo {author} {\bibfnamefont {V.~S.}\ \bibnamefont
  {de~Carvalho}}, \bibinfo {author} {\bibfnamefont {T.}~\bibnamefont {Birol}},
  \ and\ \bibinfo {author} {\bibfnamefont {R.~G.}\ \bibnamefont {Pereira}},\
  }\bibfield  {title} {\enquote {\bibinfo {title} {Topological transition from
  nodal to nodeless zeeman splitting in altermagnets}}, }\href {\doibase
  10.1103/PhysRevB.109.024404} {\bibfield  {journal} {\bibinfo  {journal}
  {Phys. Rev. B}\ }\textbf {\bibinfo {volume} {109}},\ \bibinfo {pages}
  {024404} (\bibinfo {year} {2024})}\BibitemShut {NoStop}%
\bibitem [{\citenamefont {Zhang}\ \emph {et~al.}(2024)\citenamefont {Zhang},
  \citenamefont {Cui}, \citenamefont {Li}, \citenamefont {Duan}, \citenamefont
  {Li}, \citenamefont {Yu},\ and\ \citenamefont {Yao}}]{zhang2024prl}%
  \BibitemOpen
  \bibfield  {author} {\bibinfo {author} {\bibfnamefont {R.-W.}\ \bibnamefont
  {Zhang}}, \bibinfo {author} {\bibfnamefont {C.}~\bibnamefont {Cui}}, \bibinfo
  {author} {\bibfnamefont {R.}~\bibnamefont {Li}}, \bibinfo {author}
  {\bibfnamefont {J.}~\bibnamefont {Duan}}, \bibinfo {author} {\bibfnamefont
  {L.}~\bibnamefont {Li}}, \bibinfo {author} {\bibfnamefont {Z.-M.}\
  \bibnamefont {Yu}}, \ and\ \bibinfo {author} {\bibfnamefont {Y.}~\bibnamefont
  {Yao}},\ }\bibfield  {title} {\enquote {\bibinfo {title} {Predictable
  gate-field control of spin in altermagnets with spin-layer coupling}}, }\href
  {\doibase 10.1103/PhysRevLett.133.056401} {\bibfield  {journal} {\bibinfo
  {journal} {Phys. Rev. Lett.}\ }\textbf {\bibinfo {volume} {133}},\ \bibinfo
  {pages} {056401} (\bibinfo {year} {2024})}\BibitemShut {NoStop}%
\bibitem [{\citenamefont {Zhou}\ \emph {et~al.}(2024)\citenamefont {Zhou},
  \citenamefont {Feng}, \citenamefont {Zhang}, \citenamefont
  {\ifmmode~\check{S}\else \v{S}\fi{}mejkal}, \citenamefont {Sinova},
  \citenamefont {Mokrousov},\ and\ \citenamefont {Yao}}]{Zhou2024PRL}%
  \BibitemOpen
  \bibfield  {author} {\bibinfo {author} {\bibfnamefont {X.}~\bibnamefont
  {Zhou}}, \bibinfo {author} {\bibfnamefont {W.}~\bibnamefont {Feng}}, \bibinfo
  {author} {\bibfnamefont {R.-W.}\ \bibnamefont {Zhang}}, \bibinfo {author}
  {\bibfnamefont {L.}~\bibnamefont {\ifmmode~\check{S}\else \v{S}\fi{}mejkal}},
  \bibinfo {author} {\bibfnamefont {J.}~\bibnamefont {Sinova}}, \bibinfo
  {author} {\bibfnamefont {Y.}~\bibnamefont {Mokrousov}}, \ and\ \bibinfo
  {author} {\bibfnamefont {Y.}~\bibnamefont {Yao}},\ }\bibfield  {title}
  {\enquote {\bibinfo {title} {{Crystal Thermal Transport in Altermagnetic
  ${\mathrm{RuO}}_{2}$}}}, }\href {\doibase 10.1103/PhysRevLett.132.056701}
  {\bibfield  {journal} {\bibinfo  {journal} {Phys. Rev. Lett.}\ }\textbf
  {\bibinfo {volume} {132}},\ \bibinfo {pages} {056701} (\bibinfo {year}
  {2024})}\BibitemShut {NoStop}%
\bibitem [{\citenamefont {Chen}\ \emph
  {et~al.}(2025{\natexlab{a}})\citenamefont {Chen}, \citenamefont {Liu},
  \citenamefont {Lu},\ and\ \citenamefont {Xie}}]{ChenYY2025PRL}%
  \BibitemOpen
  \bibfield  {author} {\bibinfo {author} {\bibfnamefont {Y.}~\bibnamefont
  {Chen}}, \bibinfo {author} {\bibfnamefont {X.}~\bibnamefont {Liu}}, \bibinfo
  {author} {\bibfnamefont {H.-Z.}\ \bibnamefont {Lu}}, \ and\ \bibinfo {author}
  {\bibfnamefont {X.~C.}\ \bibnamefont {Xie}},\ }\bibfield  {title} {\enquote
  {\bibinfo {title} {Electrical switching of altermagnetism}}, }\href {\doibase
  10.1103/zm5y-vy41} {\bibfield  {journal} {\bibinfo  {journal} {Phys. Rev.
  Lett.}\ }\textbf {\bibinfo {volume} {135}},\ \bibinfo {pages} {016701}
  (\bibinfo {year} {2025}{\natexlab{a}})}\BibitemShut {NoStop}%
\bibitem [{\citenamefont {Li}\ \emph {et~al.}(2024{\natexlab{a}})\citenamefont
  {Li}, \citenamefont {Wang}, \citenamefont {Ding}, \citenamefont {Ren},
  \citenamefont {Zhao}, \citenamefont {Lin}, \citenamefont {Yang},
  \citenamefont {Yan}, \citenamefont {Li}, \citenamefont {Yang}, \citenamefont
  {Yuan}, \citenamefont {Denlinger}, \citenamefont {Wang}, \citenamefont
  {Zhang}, \citenamefont {Wray}, \citenamefont {Dong}, \citenamefont {Qian},\
  and\ \citenamefont {Miao}}]{sdongFeSb2}%
  \BibitemOpen
  \bibfield  {author} {\bibinfo {author} {\bibfnamefont {H.}~\bibnamefont
  {Li}}, \bibinfo {author} {\bibfnamefont {G.}~\bibnamefont {Wang}}, \bibinfo
  {author} {\bibfnamefont {N.}~\bibnamefont {Ding}}, \bibinfo {author}
  {\bibfnamefont {Q.}~\bibnamefont {Ren}}, \bibinfo {author} {\bibfnamefont
  {G.}~\bibnamefont {Zhao}}, \bibinfo {author} {\bibfnamefont {W.}~\bibnamefont
  {Lin}},  \emph {et~al.},\ }\bibfield  {title} {\enquote {\bibinfo {title}
  {{Spectroscopic evidence of spin-state excitation in d-electron correlated
  semiconductor FeSb$_2$}}}, }\href {https://doi.org/10.1073/pnas.2321193121}
  {\bibfield  {journal} {\bibinfo  {journal} {Proc. Nat. Acad. Sci.}\ }\textbf
  {\bibinfo {volume} {121}},\ \bibinfo {pages} {e2321193121} (\bibinfo {year}
  {2024}{\natexlab{a}})}\BibitemShut {NoStop}%
\bibitem [{\citenamefont {\ifmmode~\check{S}\else \v{S}\fi{}mejkal}\ \emph
  {et~al.}(2022{\natexlab{c}})\citenamefont {\ifmmode~\check{S}\else
  \v{S}\fi{}mejkal}, \citenamefont {Hellenes}, \citenamefont
  {Gonz\'alez-Hern\'andez}, \citenamefont {Sinova},\ and\ \citenamefont
  {Jungwirth}}]{ifmmode2022Giant}%
  \BibitemOpen
  \bibfield  {author} {\bibinfo {author} {\bibfnamefont {L.}~\bibnamefont
  {\ifmmode~\check{S}\else \v{S}\fi{}mejkal}}, \bibinfo {author} {\bibfnamefont
  {A.~B.}\ \bibnamefont {Hellenes}}, \bibinfo {author} {\bibfnamefont
  {R.}~\bibnamefont {Gonz\'alez-Hern\'andez}}, \bibinfo {author} {\bibfnamefont
  {J.}~\bibnamefont {Sinova}}, \ and\ \bibinfo {author} {\bibfnamefont
  {T.}~\bibnamefont {Jungwirth}},\ }\bibfield  {title} {\enquote {\bibinfo
  {title} {Giant and tunneling magnetoresistance in unconventional collinear
  antiferromagnets with nonrelativistic spin-momentum coupling}}, }\href
  {\doibase 10.1103/PhysRevX.12.011028} {\bibfield  {journal} {\bibinfo
  {journal} {Phys. Rev. X}\ }\textbf {\bibinfo {volume} {12}},\ \bibinfo
  {pages} {011028} (\bibinfo {year} {2022}{\natexlab{c}})}\BibitemShut
  {NoStop}%
\bibitem [{\citenamefont {Ezawa}(2024)}]{Ezawa2024PRLAlter}%
  \BibitemOpen
  \bibfield  {author} {\bibinfo {author} {\bibfnamefont {M.}~\bibnamefont
  {Ezawa}},\ }\bibfield  {title} {\enquote {\bibinfo {title} {{Detecting the
  N\'eel vector of altermagnets in heterostructures with a topological
  insulator and a crystalline valley-edge insulator}}}, }\href {\doibase
  10.1103/PhysRevB.109.245306} {\bibfield  {journal} {\bibinfo  {journal}
  {Phys. Rev. B}\ }\textbf {\bibinfo {volume} {109}},\ \bibinfo {pages}
  {245306} (\bibinfo {year} {2024})}\BibitemShut {NoStop}%
\bibitem [{\citenamefont {Li}\ \emph {et~al.}(2024{\natexlab{b}})\citenamefont
  {Li}, \citenamefont {Liu},\ and\ \citenamefont {Liu}}]{LiuCC2024PRB}%
  \BibitemOpen
  \bibfield  {author} {\bibinfo {author} {\bibfnamefont {Y.-X.}\ \bibnamefont
  {Li}}, \bibinfo {author} {\bibfnamefont {Y.}~\bibnamefont {Liu}}, \ and\
  \bibinfo {author} {\bibfnamefont {C.-C.}\ \bibnamefont {Liu}},\ }\bibfield
  {title} {\enquote {\bibinfo {title} {Creation and manipulation of
  higher-order topological states by altermagnets}}, }\href {\doibase
  10.1103/PhysRevB.109.L201109} {\bibfield  {journal} {\bibinfo  {journal}
  {Phys. Rev. B}\ }\textbf {\bibinfo {volume} {109}},\ \bibinfo {pages}
  {L201109} (\bibinfo {year} {2024}{\natexlab{b}})}\BibitemShut {NoStop}%
\bibitem [{\citenamefont {Chen}\ \emph
  {et~al.}(2025{\natexlab{b}})\citenamefont {Chen}, \citenamefont {Yi},
  \citenamefont {Zhou},\ and\ \citenamefont {Xu}}]{ChenR2024altermagnet}%
  \BibitemOpen
  \bibfield  {author} {\bibinfo {author} {\bibfnamefont {R.}~\bibnamefont
  {Chen}}, \bibinfo {author} {\bibfnamefont {X.-X.}\ \bibnamefont {Yi}},
  \bibinfo {author} {\bibfnamefont {B.}~\bibnamefont {Zhou}}, \ and\ \bibinfo
  {author} {\bibfnamefont {D.-H.}\ \bibnamefont {Xu}},\ }\bibfield  {title}
  {\enquote {\bibinfo {title} {Anomalous hall effects in magnetic weak
  topological insulator films}}, }\href {\doibase 10.1103/PhysRevB.111.045409}
  {\bibfield  {journal} {\bibinfo  {journal} {Phys. Rev. B}\ }\textbf {\bibinfo
  {volume} {111}},\ \bibinfo {pages} {045409} (\bibinfo {year}
  {2025}{\natexlab{b}})}\BibitemShut {NoStop}%
\bibitem [{\citenamefont {Chen}\ \emph
  {et~al.}(2025{\natexlab{c}})\citenamefont {Chen}, \citenamefont {Wang},
  \citenamefont {Sun}, \citenamefont {Zhou},\ and\ \citenamefont
  {Xu}}]{ChenR2025altermagnet}%
  \BibitemOpen
  \bibfield  {author} {\bibinfo {author} {\bibfnamefont {R.}~\bibnamefont
  {Chen}}, \bibinfo {author} {\bibfnamefont {Z.-M.}\ \bibnamefont {Wang}},
  \bibinfo {author} {\bibfnamefont {H.-P.}\ \bibnamefont {Sun}}, \bibinfo
  {author} {\bibfnamefont {B.}~\bibnamefont {Zhou}}, \ and\ \bibinfo {author}
  {\bibfnamefont {D.-H.}\ \bibnamefont {Xu}},\ }\bibfield  {title} {\enquote
  {\bibinfo {title} {Probing $ k $-space alternating spin polarization via the
  anomalous hall effect}}, }\href {https://arxiv.org/pdf/2501.14217} {\bibfield
   {journal} {\bibinfo  {journal} {arXiv:2501.14217}\ } (\bibinfo {year}
  {2025}{\natexlab{c}})}\BibitemShut {NoStop}%
\bibitem [{\citenamefont {Bai}\ \emph {et~al.}(2023)\citenamefont {Bai},
  \citenamefont {Zhang}, \citenamefont {Zhou}, \citenamefont {Chen},
  \citenamefont {Wan}, \citenamefont {Han}, \citenamefont {Zhu}, \citenamefont
  {Liang}, \citenamefont {Su}, \citenamefont {Han}, \citenamefont {Pan},\ and\
  \citenamefont {Song}}]{Bai2023PRL}%
  \BibitemOpen
  \bibfield  {author} {\bibinfo {author} {\bibfnamefont {H.}~\bibnamefont
  {Bai}}, \bibinfo {author} {\bibfnamefont {Y.~C.}\ \bibnamefont {Zhang}},
  \bibinfo {author} {\bibfnamefont {Y.~J.}\ \bibnamefont {Zhou}}, \bibinfo
  {author} {\bibfnamefont {P.}~\bibnamefont {Chen}}, \bibinfo {author}
  {\bibfnamefont {C.~H.}\ \bibnamefont {Wan}}, \bibinfo {author} {\bibfnamefont
  {L.}~\bibnamefont {Han}},  \emph {et~al.},\ }\bibfield  {title} {\enquote
  {\bibinfo {title} {Efficient spin-to-charge conversion via altermagnetic spin
  splitting effect in antiferromagnet ${\mathrm{ruo}}_{2}$}}, }\href {\doibase
  10.1103/PhysRevLett.130.216701} {\bibfield  {journal} {\bibinfo  {journal}
  {Phys. Rev. Lett.}\ }\textbf {\bibinfo {volume} {130}},\ \bibinfo {pages}
  {216701} (\bibinfo {year} {2023})}\BibitemShut {NoStop}%
\bibitem [{\citenamefont {Gonz\'alez-Hern\'andez}\ \emph
  {et~al.}(2021)\citenamefont {Gonz\'alez-Hern\'andez}, \citenamefont
  {\ifmmode~\check{S}\else \v{S}\fi{}mejkal}, \citenamefont {V\'yborn\'y},
  \citenamefont {Yahagi}, \citenamefont {Sinova}, \citenamefont {Jungwirth},\
  and\ \citenamefont {\ifmmode~\check{Z}\else
  \v{Z}\fi{}elezn\'y}}]{Gonzalez2021PRL}%
  \BibitemOpen
  \bibfield  {author} {\bibinfo {author} {\bibfnamefont {R.}~\bibnamefont
  {Gonz\'alez-Hern\'andez}}, \bibinfo {author} {\bibfnamefont {L.}~\bibnamefont
  {\ifmmode~\check{S}\else \v{S}\fi{}mejkal}}, \bibinfo {author} {\bibfnamefont
  {K.}~\bibnamefont {V\'yborn\'y}}, \bibinfo {author} {\bibfnamefont
  {Y.}~\bibnamefont {Yahagi}}, \bibinfo {author} {\bibfnamefont
  {J.}~\bibnamefont {Sinova}}, \bibinfo {author} {\bibfnamefont {T.~c.~v.}\
  \bibnamefont {Jungwirth}}, \ and\ \bibinfo {author} {\bibfnamefont
  {J.}~\bibnamefont {\ifmmode~\check{Z}\else \v{Z}\fi{}elezn\'y}},\ }\bibfield
  {title} {\enquote {\bibinfo {title} {Efficient electrical spin splitter based
  on nonrelativistic collinear antiferromagnetism}}, }\href {\doibase
  10.1103/PhysRevLett.126.127701} {\bibfield  {journal} {\bibinfo  {journal}
  {Phys. Rev. Lett.}\ }\textbf {\bibinfo {volume} {126}},\ \bibinfo {pages}
  {127701} (\bibinfo {year} {2021})}\BibitemShut {NoStop}%
\bibitem [{\citenamefont {{\v{S}}mejkal}\ \emph {et~al.}(2020)\citenamefont
  {{\v{S}}mejkal}, \citenamefont {Gonz{\'a}lez-Hern{\'a}ndez}, \citenamefont
  {Jungwirth},\ and\ \citenamefont {Sinova}}]{vsmejkal2020crystal}%
  \BibitemOpen
  \bibfield  {author} {\bibinfo {author} {\bibfnamefont {L.}~\bibnamefont
  {{\v{S}}mejkal}}, \bibinfo {author} {\bibfnamefont {R.}~\bibnamefont
  {Gonz{\'a}lez-Hern{\'a}ndez}}, \bibinfo {author} {\bibfnamefont
  {T.}~\bibnamefont {Jungwirth}}, \ and\ \bibinfo {author} {\bibfnamefont
  {J.}~\bibnamefont {Sinova}},\ }\bibfield  {title} {\enquote {\bibinfo {title}
  {{Crystal time-reversal symmetry breaking and spontaneous Hall effect in
  collinear antiferromagnets}}}, }\href
  {https://www.science.org/doi/full/10.1126/sciadv.aaz8809} {\bibfield
  {journal} {\bibinfo  {journal} {Sci. Adv.}\ }\textbf {\bibinfo {volume}
  {6}},\ \bibinfo {pages} {eaaz8809} (\bibinfo {year} {2020})}\BibitemShut
  {NoStop}%
\bibitem [{\citenamefont {Feng}\ \emph {et~al.}(2022)\citenamefont {Feng},
  \citenamefont {Zhou}, \citenamefont {Smejkal}, \citenamefont {Wu},
  \citenamefont {Zhu}, \citenamefont {Guo}, \citenamefont
  {González-Hernández}, \citenamefont {Wang}, \citenamefont {Yan},
  \citenamefont {Qin}, \citenamefont {Zhang}, \citenamefont {Wu}, \citenamefont
  {Chen}, \citenamefont {Meng}, \citenamefont {Liu}, \citenamefont {Xia},
  \citenamefont {Sinova}, \citenamefont {Jungwirth},\ and\ \citenamefont
  {Liu}}]{Feng2022NatElec}%
  \BibitemOpen
  \bibfield  {author} {\bibinfo {author} {\bibfnamefont {Z.}~\bibnamefont
  {Feng}}, \bibinfo {author} {\bibfnamefont {X.}~\bibnamefont {Zhou}}, \bibinfo
  {author} {\bibfnamefont {L.}~\bibnamefont {Smejkal}}, \bibinfo {author}
  {\bibfnamefont {L.}~\bibnamefont {Wu}}, \bibinfo {author} {\bibfnamefont
  {Z.}~\bibnamefont {Zhu}}, \bibinfo {author} {\bibfnamefont {H.}~\bibnamefont
  {Guo}},  \emph {et~al.},\ }\bibfield  {title} {\enquote {\bibinfo {title} {An
  anomalous {Hall} effect in altermagnetic ruthenium dioxide}}, }\href
  {\doibase 10.1038/s41928-022-00866-z} {\bibfield  {journal} {\bibinfo
  {journal} {Nat. Elec.}\ }\textbf {\bibinfo {volume} {5}},\ \bibinfo {pages}
  {735} (\bibinfo {year} {2022})}\BibitemShut {NoStop}%
\bibitem [{\citenamefont {Fedchenko}\ \emph {et~al.}(2024)\citenamefont
  {Fedchenko}, \citenamefont {Minar}, \citenamefont {Akashdeep}, \citenamefont
  {DSouza}, \citenamefont {Vasilyev}, \citenamefont {Tkach}, \citenamefont
  {Odenbreit}, \citenamefont {Nguyen}, \citenamefont {Kutnyakhov},
  \citenamefont {Wind}, \citenamefont {Wenthaus}, \citenamefont {Scholz},
  \citenamefont {Rossnagel}, \citenamefont {Hoesch}, \citenamefont
  {Aeschlimann}, \citenamefont {Stadtmüller}, \citenamefont {Klaui},
  \citenamefont {Schönhense}, \citenamefont {Jungwirth}, \citenamefont
  {Hellenes}, \citenamefont {Jakob}, \citenamefont {Smejkal}, \citenamefont
  {Sinova},\ and\ \citenamefont {Elmers}}]{Fedchenko2024SciAdv}%
  \BibitemOpen
  \bibfield  {author} {\bibinfo {author} {\bibfnamefont {O.}~\bibnamefont
  {Fedchenko}}, \bibinfo {author} {\bibfnamefont {J.}~\bibnamefont {Minar}},
  \bibinfo {author} {\bibfnamefont {A.}~\bibnamefont {Akashdeep}}, \bibinfo
  {author} {\bibfnamefont {S.~W.}\ \bibnamefont {DSouza}}, \bibinfo {author}
  {\bibfnamefont {D.}~\bibnamefont {Vasilyev}}, \bibinfo {author}
  {\bibfnamefont {O.}~\bibnamefont {Tkach}},  \emph {et~al.},\ }\bibfield
  {title} {\enquote {\bibinfo {title} {{Observation of time-reversal symmetry
  breaking in the band structure of altermagnetic RuO$_2$}}}, }\href {\doibase
  10.1126/sciadv.adj4883} {\bibfield  {journal} {\bibinfo  {journal} {Sci.
  Adv.}\ }\textbf {\bibinfo {volume} {10}},\ \bibinfo {pages} {eadj4883}
  (\bibinfo {year} {2024})}\BibitemShut {NoStop}%
\bibitem [{\citenamefont {Karube}\ \emph {et~al.}(2022)\citenamefont {Karube},
  \citenamefont {Tanaka}, \citenamefont {Sugawara}, \citenamefont {Kadoguchi},
  \citenamefont {Kohda},\ and\ \citenamefont {Nitta}}]{Karube2022PRL}%
  \BibitemOpen
  \bibfield  {author} {\bibinfo {author} {\bibfnamefont {S.}~\bibnamefont
  {Karube}}, \bibinfo {author} {\bibfnamefont {T.}~\bibnamefont {Tanaka}},
  \bibinfo {author} {\bibfnamefont {D.}~\bibnamefont {Sugawara}}, \bibinfo
  {author} {\bibfnamefont {N.}~\bibnamefont {Kadoguchi}}, \bibinfo {author}
  {\bibfnamefont {M.}~\bibnamefont {Kohda}}, \ and\ \bibinfo {author}
  {\bibfnamefont {J.}~\bibnamefont {Nitta}},\ }\bibfield  {title} {\enquote
  {\bibinfo {title} {{Observation of Spin-Splitter Torque in Collinear
  Antiferromagnetic ${\mathrm{RuO}}_{2}$}}}, }\href {\doibase
  10.1103/PhysRevLett.129.137201} {\bibfield  {journal} {\bibinfo  {journal}
  {Phys. Rev. Lett.}\ }\textbf {\bibinfo {volume} {129}},\ \bibinfo {pages}
  {137201} (\bibinfo {year} {2022})}\BibitemShut {NoStop}%
\bibitem [{\citenamefont {Ke{\ss}ler}\ \emph {et~al.}(2024)\citenamefont
  {Ke{\ss}ler}, \citenamefont {Garcia-Gassull}, \citenamefont {Suter},
  \citenamefont {Prokscha}, \citenamefont {Salman}, \citenamefont {Khalyavin},
  \citenamefont {Manuel}, \citenamefont {Orlandi}, \citenamefont {Mazin},
  \citenamefont {Valentí},\ and\ \citenamefont {Moser}}]{Keler2024NpjS}%
  \BibitemOpen
  \bibfield  {author} {\bibinfo {author} {\bibfnamefont {P.}~\bibnamefont
  {Ke{\ss}ler}}, \bibinfo {author} {\bibfnamefont {L.}~\bibnamefont
  {Garcia-Gassull}}, \bibinfo {author} {\bibfnamefont {A.}~\bibnamefont
  {Suter}}, \bibinfo {author} {\bibfnamefont {T.}~\bibnamefont {Prokscha}},
  \bibinfo {author} {\bibfnamefont {Z.}~\bibnamefont {Salman}}, \bibinfo
  {author} {\bibfnamefont {D.}~\bibnamefont {Khalyavin}},  \emph {et~al.},\
  }\bibfield  {title} {\enquote {\bibinfo {title} {{Absence of magnetic order
  in RuO$_2$: insights from $\mu$SR spectroscopy and neutron diffraction}}},
  }\href {\doibase 10.1038/s44306-024-00055-y} {\bibfield  {journal} {\bibinfo
  {journal} {npj Spintronics}\ }\textbf {\bibinfo {volume} {2}},\ \bibinfo
  {pages} {50} (\bibinfo {year} {2024})}\BibitemShut {NoStop}%
\bibitem [{\citenamefont {Leeb}\ \emph {et~al.}(2026)\citenamefont {Leeb},
  \citenamefont {d'Ornellas}, \citenamefont {de~Juan},\ and\ \citenamefont
  {Grushin}}]{Leeb2026arXiv}%
  \BibitemOpen
  \bibfield  {author} {\bibinfo {author} {\bibfnamefont {V.}~\bibnamefont
  {Leeb}}, \bibinfo {author} {\bibfnamefont {P.}~\bibnamefont {d'Ornellas}},
  \bibinfo {author} {\bibfnamefont {F.}~\bibnamefont {de~Juan}}, \ and\
  \bibinfo {author} {\bibfnamefont {A.~G.}\ \bibnamefont {Grushin}},\
  }\bibfield  {title} {\enquote {\bibinfo {title} {Topologically protected
  surface altermagnetism on antiferromagnets}}, }\href
  {https://arxiv.org/pdf/2602.10108} {\bibfield  {journal} {\bibinfo  {journal}
  {arXiv:2602.10108}\ } (\bibinfo {year} {2026})}\BibitemShut {NoStop}%
\bibitem [{\citenamefont {Lange}\ \emph {et~al.}(2026)\citenamefont {Lange},
  \citenamefont {Jaeschke-Ubiergo}, \citenamefont {Chakraborty}, \citenamefont
  {Verbeek}, \citenamefont {{\v{S}}mejkal}, \citenamefont {Sinova},\ and\
  \citenamefont {Mook}}]{Lange2026arXiv}%
  \BibitemOpen
  \bibfield  {author} {\bibinfo {author} {\bibfnamefont {C.}~\bibnamefont
  {Lange}}, \bibinfo {author} {\bibfnamefont {R.}~\bibnamefont
  {Jaeschke-Ubiergo}}, \bibinfo {author} {\bibfnamefont {A.}~\bibnamefont
  {Chakraborty}}, \bibinfo {author} {\bibfnamefont {X.~H.}\ \bibnamefont
  {Verbeek}}, \bibinfo {author} {\bibfnamefont {L.}~\bibnamefont
  {{\v{S}}mejkal}}, \bibinfo {author} {\bibfnamefont {J.}~\bibnamefont
  {Sinova}}, \ and\ \bibinfo {author} {\bibfnamefont {A.}~\bibnamefont
  {Mook}},\ }\bibfield  {title} {\enquote {\bibinfo {title} {Emergent
  altermagnetism at surfaces of antiferromagnets: full symmetry classification
  and material identification}}, }\href {https://arxiv.org/pdf/2602.08773}
  {\bibfield  {journal} {\bibinfo  {journal} {arXiv:2602.08773}\ } (\bibinfo
  {year} {2026})}\BibitemShut {NoStop}%
\bibitem [{\citenamefont {Sasioglu}\ \emph {et~al.}(2026)\citenamefont
  {Sasioglu}, \citenamefont {Mertig},\ and\ \citenamefont
  {Lounis}}]{Sasioglu2026arXiv}%
  \BibitemOpen
  \bibfield  {author} {\bibinfo {author} {\bibfnamefont {E.}~\bibnamefont
  {Sasioglu}}, \bibinfo {author} {\bibfnamefont {I.}~\bibnamefont {Mertig}}, \
  and\ \bibinfo {author} {\bibfnamefont {S.}~\bibnamefont {Lounis}},\
  }\bibfield  {title} {\enquote {\bibinfo {title} {$d$-wave surface
  altermagnetism in centrosymmetric collinear antiferromagnets}}, }\href
  {https://arxiv.org/pdf/2602.08790} {\bibfield  {journal} {\bibinfo  {journal}
  {arXiv:2602.08790}\ } (\bibinfo {year} {2026})}\BibitemShut {NoStop}%
\bibitem [{\citenamefont {Hu}\ \emph {et~al.}(2026{\natexlab{a}})\citenamefont
  {Hu}, \citenamefont {Zhou}, \citenamefont {Pan}, \citenamefont {Liu},
  \citenamefont {Zhou},\ and\ \citenamefont {Sun}}]{HuYZ2026PRL}%
  \BibitemOpen
  \bibfield  {author} {\bibinfo {author} {\bibfnamefont {Y.}~\bibnamefont
  {Hu}}, \bibinfo {author} {\bibfnamefont {P.}~\bibnamefont {Zhou}}, \bibinfo
  {author} {\bibfnamefont {B.}~\bibnamefont {Pan}}, \bibinfo {author}
  {\bibfnamefont {S.}~\bibnamefont {Liu}}, \bibinfo {author} {\bibfnamefont
  {B.}~\bibnamefont {Zhou}}, \ and\ \bibinfo {author} {\bibfnamefont
  {L.}~\bibnamefont {Sun}},\ }\bibfield  {title} {\enquote {\bibinfo {title}
  {Emergent surface altermagnetism}}, }\href {\doibase 10.1103/hw4l-jknk}
  {\bibfield  {journal} {\bibinfo  {journal} {Phys. Rev. Lett.}\ } (\bibinfo
  {year} {2026}{\natexlab{a}}),\ 10.1103/hw4l-jknk}\BibitemShut {NoStop}%
\bibitem [{\citenamefont {Su}\ \emph {et~al.}(1979)\citenamefont {Su},
  \citenamefont {Schrieffer},\ and\ \citenamefont {Heeger}}]{Su1979PRL}%
  \BibitemOpen
  \bibfield  {author} {\bibinfo {author} {\bibfnamefont {W.~P.}\ \bibnamefont
  {Su}}, \bibinfo {author} {\bibfnamefont {J.~R.}\ \bibnamefont {Schrieffer}},
  \ and\ \bibinfo {author} {\bibfnamefont {A.~J.}\ \bibnamefont {Heeger}},\
  }\bibfield  {title} {\enquote {\bibinfo {title} {Solitons in polyacetylene}},
  }\href {\doibase 10.1103/PhysRevLett.42.1698} {\bibfield  {journal} {\bibinfo
   {journal} {Phys. Rev. Lett.}\ }\textbf {\bibinfo {volume} {42}},\ \bibinfo
  {pages} {1698} (\bibinfo {year} {1979})}\BibitemShut {NoStop}%
\bibitem [{\citenamefont {Guo}(2026)}]{GuoSD2026FronPhys}%
  \BibitemOpen
  \bibfield  {author} {\bibinfo {author} {\bibfnamefont {S.-D.}\ \bibnamefont
  {Guo}},\ }\bibfield  {title} {\enquote {\bibinfo {title} {Hidden
  altermagnetism}}, }\href {\doibase
  https://doi.org/10.15302/frontphys.2026.025201} {\bibfield  {journal}
  {\bibinfo  {journal} {Frontiers of Physics}\ }\textbf {\bibinfo {volume}
  {21}},\ \bibinfo {pages} {025201} (\bibinfo {year} {2026})}\BibitemShut
  {NoStop}%
\bibitem [{\citenamefont {Matsuda}\ \emph {et~al.}(2025)\citenamefont
  {Matsuda}, \citenamefont {Watanabe},\ and\ \citenamefont
  {Arita}}]{Matsuda2025PRL}%
  \BibitemOpen
  \bibfield  {author} {\bibinfo {author} {\bibfnamefont {J.}~\bibnamefont
  {Matsuda}}, \bibinfo {author} {\bibfnamefont {H.}~\bibnamefont {Watanabe}}, \
  and\ \bibinfo {author} {\bibfnamefont {R.}~\bibnamefont {Arita}},\ }\bibfield
   {title} {\enquote {\bibinfo {title} {Multiferroic collinear antiferromagnets
  with hidden altermagnetic spin splitting}}, }\href {\doibase
  10.1103/vgcs-bn8g} {\bibfield  {journal} {\bibinfo  {journal} {Phys. Rev.
  Lett.}\ }\textbf {\bibinfo {volume} {134}},\ \bibinfo {pages} {226703}
  (\bibinfo {year} {2025})}\BibitemShut {NoStop}%
\bibitem [{\citenamefont {Yang}\ \emph {et~al.}(2025)\citenamefont {Yang},
  \citenamefont {Chen}, \citenamefont {Liu}, \citenamefont {Li}, \citenamefont
  {Pan}, \citenamefont {Deng}, \citenamefont {Zheng}, \citenamefont {Tang},
  \citenamefont {Zheng}, \citenamefont {Zhu} \emph {et~al.}}]{Yang2025arXiv}%
  \BibitemOpen
  \bibfield  {author} {\bibinfo {author} {\bibfnamefont {G.}~\bibnamefont
  {Yang}}, \bibinfo {author} {\bibfnamefont {R.}~\bibnamefont {Chen}}, \bibinfo
  {author} {\bibfnamefont {C.}~\bibnamefont {Liu}}, \bibinfo {author}
  {\bibfnamefont {J.}~\bibnamefont {Li}}, \bibinfo {author} {\bibfnamefont
  {Z.}~\bibnamefont {Pan}}, \bibinfo {author} {\bibfnamefont {L.}~\bibnamefont
  {Deng}},  \emph {et~al.},\ }\bibfield  {title} {\enquote {\bibinfo {title}
  {{Observation of hidden altermagnetism in Cs$_{1-\delta}$V$_2$Te$_2$O}}},
  }\href {https://arxiv.org/pdf/2512.00972} {\bibfield  {journal} {\bibinfo
  {journal} {arXiv:2512.00972}\ } (\bibinfo {year} {2025})}\BibitemShut
  {NoStop}%
\bibitem [{\citenamefont {Li}\ \emph {et~al.}(2026{\natexlab{b}})\citenamefont
  {Li}, \citenamefont {Hou}, \citenamefont {Zhu}, \citenamefont {Zheng},
  \citenamefont {Song}, \citenamefont {Liu}, \citenamefont {Zhang},\ and\
  \citenamefont {Hu}}]{LiChuang2026arXiv}%
  \BibitemOpen
  \bibfield  {author} {\bibinfo {author} {\bibfnamefont {C.}~\bibnamefont
  {Li}}, \bibinfo {author} {\bibfnamefont {J.-X.}\ \bibnamefont {Hou}},
  \bibinfo {author} {\bibfnamefont {S.-L.}\ \bibnamefont {Zhu}}, \bibinfo
  {author} {\bibfnamefont {H.}~\bibnamefont {Zheng}}, \bibinfo {author}
  {\bibfnamefont {Y.}~\bibnamefont {Song}}, \bibinfo {author} {\bibfnamefont
  {Y.}~\bibnamefont {Liu}}, \bibinfo {author} {\bibfnamefont {S.-B.}\
  \bibnamefont {Zhang}}, \ and\ \bibinfo {author} {\bibfnamefont {L.-H.}\
  \bibnamefont {Hu}},\ }\bibfield  {title} {\enquote {\bibinfo {title}
  {Altermagnetic even-odd effects in {CsV$_2$Te$_2$O} josephson junctions}},
  }\href {https://arxiv.org/pdf/2602.14485} {\bibfield  {journal} {\bibinfo
  {journal} {arXiv:2602.14485}\ } (\bibinfo {year}
  {2026}{\natexlab{b}})}\BibitemShut {NoStop}%
\bibitem [{\citenamefont {Hu}\ \emph {et~al.}(2026{\natexlab{b}})\citenamefont
  {Hu}, \citenamefont {Zhou}, \citenamefont {Pan}, \citenamefont {Lyu},\ and\
  \citenamefont {Sun}}]{HuYZ2026PRL1}%
  \BibitemOpen
  \bibfield  {author} {\bibinfo {author} {\bibfnamefont {Y.}~\bibnamefont
  {Hu}}, \bibinfo {author} {\bibfnamefont {P.}~\bibnamefont {Zhou}}, \bibinfo
  {author} {\bibfnamefont {B.}~\bibnamefont {Pan}}, \bibinfo {author}
  {\bibfnamefont {P.}~\bibnamefont {Lyu}}, \ and\ \bibinfo {author}
  {\bibfnamefont {L.}~\bibnamefont {Sun}},\ }\bibfield  {title} {\enquote
  {\bibinfo {title} {Symmetry classification of nonrelativistic hidden spin
  polarization in noncollinear magnets}}, }\href {\doibase 10.1103/8pg5-pz4z}
  {\bibfield  {journal} {\bibinfo  {journal} {Phys. Rev. Lett.}\ } (\bibinfo
  {year} {2026}{\natexlab{b}}),\ 10.1103/8pg5-pz4z}\BibitemShut {NoStop}%
\bibitem [{\citenamefont {Parkin}\ \emph {et~al.}(1990)\citenamefont {Parkin},
  \citenamefont {More},\ and\ \citenamefont {Roche}}]{Parkin1990PRL}%
  \BibitemOpen
  \bibfield  {author} {\bibinfo {author} {\bibfnamefont {S.~S.~P.}\
  \bibnamefont {Parkin}}, \bibinfo {author} {\bibfnamefont {N.}~\bibnamefont
  {More}}, \ and\ \bibinfo {author} {\bibfnamefont {K.~P.}\ \bibnamefont
  {Roche}},\ }\bibfield  {title} {\enquote {\bibinfo {title} {{Oscillations in
  exchange coupling and magnetoresistance in metallic superlattice structures:
  Co/Ru, Co/Cr, and Fe/Cr}}}, }\href {\doibase 10.1103/PhysRevLett.64.2304}
  {\bibfield  {journal} {\bibinfo  {journal} {Phys. Rev. Lett.}\ }\textbf
  {\bibinfo {volume} {64}},\ \bibinfo {pages} {2304} (\bibinfo {year}
  {1990})}\BibitemShut {NoStop}%
\bibitem [{\citenamefont {Geim}\ and\ \citenamefont
  {Grigorieva}(2013)}]{Geim2013Nature}%
  \BibitemOpen
  \bibfield  {author} {\bibinfo {author} {\bibfnamefont {A.~K.}\ \bibnamefont
  {Geim}}\ and\ \bibinfo {author} {\bibfnamefont {I.~V.}\ \bibnamefont
  {Grigorieva}},\ }\bibfield  {title} {\enquote {\bibinfo {title} {Van der
  waals heterostructures}}, }\href {\doibase 10.1038/nature12385} {\bibfield
  {journal} {\bibinfo  {journal} {Nature}\ }\textbf {\bibinfo {volume} {499}},\
  \bibinfo {pages} {419–425} (\bibinfo {year} {2013})}\BibitemShut {NoStop}%
\bibitem [{\citenamefont {Song}\ \emph {et~al.}(2018)\citenamefont {Song},
  \citenamefont {Cai}, \citenamefont {Tu}, \citenamefont {Zhang}, \citenamefont
  {Huang}, \citenamefont {Wilson}, \citenamefont {Seyler}, \citenamefont {Zhu},
  \citenamefont {Taniguchi}, \citenamefont {Watanabe}, \citenamefont {McGuire},
  \citenamefont {Cobden}, \citenamefont {Xiao}, \citenamefont {Yao},\ and\
  \citenamefont {Xu}}]{SongTC2018Science}%
  \BibitemOpen
  \bibfield  {author} {\bibinfo {author} {\bibfnamefont {T.}~\bibnamefont
  {Song}}, \bibinfo {author} {\bibfnamefont {X.}~\bibnamefont {Cai}}, \bibinfo
  {author} {\bibfnamefont {M.~W.-Y.}\ \bibnamefont {Tu}}, \bibinfo {author}
  {\bibfnamefont {X.}~\bibnamefont {Zhang}}, \bibinfo {author} {\bibfnamefont
  {B.}~\bibnamefont {Huang}}, \bibinfo {author} {\bibfnamefont {N.~P.}\
  \bibnamefont {Wilson}},  \emph {et~al.},\ }\bibfield  {title} {\enquote
  {\bibinfo {title} {Giant tunneling magnetoresistance in spin-filter van der
  waals heterostructures}}, }\href {\doibase 10.1126/science.aar4851}
  {\bibfield  {journal} {\bibinfo  {journal} {Science}\ }\textbf {\bibinfo
  {volume} {360}},\ \bibinfo {pages} {1214–1218} (\bibinfo {year}
  {2018})}\BibitemShut {NoStop}%
\bibitem [{\citenamefont {Gibertini}\ \emph {et~al.}(2019)\citenamefont
  {Gibertini}, \citenamefont {Koperski}, \citenamefont {Morpurgo},\ and\
  \citenamefont {Novoselov}}]{Gibertini2019NatureNat}%
  \BibitemOpen
  \bibfield  {author} {\bibinfo {author} {\bibfnamefont {M.}~\bibnamefont
  {Gibertini}}, \bibinfo {author} {\bibfnamefont {M.}~\bibnamefont {Koperski}},
  \bibinfo {author} {\bibfnamefont {A.~F.}\ \bibnamefont {Morpurgo}}, \ and\
  \bibinfo {author} {\bibfnamefont {K.~S.}\ \bibnamefont {Novoselov}},\
  }\bibfield  {title} {\enquote {\bibinfo {title} {Magnetic 2d materials and
  heterostructures}}, }\href {\doibase 10.1038/s41565-019-0438-6} {\bibfield
  {journal} {\bibinfo  {journal} {Nature Nanotechnology}\ }\textbf {\bibinfo
  {volume} {14}},\ \bibinfo {pages} {408–419} (\bibinfo {year}
  {2019})}\BibitemShut {NoStop}%
\bibitem [{\citenamefont {Shen}(2017)}]{Shen2017TI}%
  \BibitemOpen
  \bibfield  {author} {\bibinfo {author} {\bibfnamefont {S.-Q.}\ \bibnamefont
  {Shen}},\ }\href {\doibase 10.1007/978-981-10-4606-3} {\emph {\bibinfo
  {title} {Topological Insulators--Dirac Equation in Condensed Matter}}}\
  (\bibinfo  {publisher} {Springer Singapore},\ \bibinfo {year}
  {2017})\BibitemShut {NoStop}%
\bibitem [{\citenamefont {Asbóth}\ \emph {et~al.}(2016)\citenamefont
  {Asbóth}, \citenamefont {Oroszlány},\ and\ \citenamefont
  {Pályi}}]{Asbth2016Note}%
  \BibitemOpen
  \bibfield  {author} {\bibinfo {author} {\bibfnamefont {J.~K.}\ \bibnamefont
  {Asbóth}}, \bibinfo {author} {\bibfnamefont {L.}~\bibnamefont {Oroszlány}},
  \ and\ \bibinfo {author} {\bibfnamefont {A.}~\bibnamefont {Pályi}},\
  }\enquote {\bibinfo {title} {The su-schrieffer-heeger (ssh) model}}, in\
  \href {\doibase 10.1007/978-3-319-25607-8_1} {\emph {\bibinfo {booktitle} {A
  Short Course on Topological Insulators}}}\ (\bibinfo  {publisher} {Springer
  International Publishing},\ \bibinfo {year} {2016})\ p.\ \bibinfo {pages}
  {1–22}\BibitemShut {NoStop}%
\bibitem [{Sup()}]{Supp}%
  \BibitemOpen
  \href@noop {} {\bibinfo  {journal} {See Supplemental Material for more
  details}\ }\BibitemShut {NoStop}%
\bibitem [{\citenamefont {Bernevig}\ and\ \citenamefont
  {Hughes}(2013)}]{Bernevig2013TI}%
  \BibitemOpen
\bibfield  {journal} {  }\bibfield  {author} {\bibinfo {author} {\bibfnamefont
  {B.~A.}\ \bibnamefont {Bernevig}}\ and\ \bibinfo {author} {\bibfnamefont
  {T.~L.}\ \bibnamefont {Hughes}},\ }\href {\doibase 10.1515/9781400846733}
  {\emph {\bibinfo {title} {Topological Insulators and Topological
  Superconductors}}}\ (\bibinfo  {publisher} {Princeton University Press},\
  \bibinfo {year} {2013})\BibitemShut {NoStop}%
\bibitem [{\citenamefont {Hasan}\ and\ \citenamefont
  {Kane}(2010)}]{Hasan2010RMP}%
  \BibitemOpen
  \bibfield  {author} {\bibinfo {author} {\bibfnamefont {M.~Z.}\ \bibnamefont
  {Hasan}}\ and\ \bibinfo {author} {\bibfnamefont {C.~L.}\ \bibnamefont
  {Kane}},\ }\bibfield  {title} {\enquote {\bibinfo {title} {{Colloquium:
  Topological insulators}}}, }\href {\doibase 10.1103/revmodphys.82.3045}
  {\bibfield  {journal} {\bibinfo  {journal} {Rev. Mod. Phys.}\ }\textbf
  {\bibinfo {volume} {82}},\ \bibinfo {pages} {3045} (\bibinfo {year}
  {2010})}\BibitemShut {NoStop}%
\bibitem [{\citenamefont {Qi}\ and\ \citenamefont {Zhang}(2011)}]{Qi2011RMP}%
  \BibitemOpen
  \bibfield  {author} {\bibinfo {author} {\bibfnamefont {X.-L.}\ \bibnamefont
  {Qi}}\ and\ \bibinfo {author} {\bibfnamefont {S.-C.}\ \bibnamefont {Zhang}},\
  }\bibfield  {title} {\enquote {\bibinfo {title} {{Topological insulators and
  superconductors}}}, }\href {\doibase 10.1103/revmodphys.83.1057} {\bibfield
  {journal} {\bibinfo  {journal} {Rev. Mod. Phys.}\ }\textbf {\bibinfo {volume}
  {83}},\ \bibinfo {pages} {1057} (\bibinfo {year} {2011})}\BibitemShut
  {NoStop}%
\bibitem [{\citenamefont {Gao}\ \emph {et~al.}(2021)\citenamefont {Gao},
  \citenamefont {Liu}, \citenamefont {Hu}, \citenamefont {Qiu}, \citenamefont
  {Tzschaschel}, \citenamefont {Ghosh}, \citenamefont {Ho}, \citenamefont
  {B{\'{e}}rub{\'{e}}}, \citenamefont {Chen}, \citenamefont {Sun},
  \citenamefont {Zhang}, \citenamefont {Zhang}, \citenamefont {Wang},
  \citenamefont {Wang}, \citenamefont {Huang}, \citenamefont {Felser},
  \citenamefont {Agarwal}, \citenamefont {Ding}, \citenamefont {Tien},
  \citenamefont {Akey}, \citenamefont {Gardener}, \citenamefont {Singh},
  \citenamefont {Watanabe}, \citenamefont {Taniguchi}, \citenamefont {Burch},
  \citenamefont {Bell}, \citenamefont {Zhou}, \citenamefont {Gao},
  \citenamefont {Lu}, \citenamefont {Bansil}, \citenamefont {Lin},
  \citenamefont {Chang}, \citenamefont {Fu}, \citenamefont {Ma}, \citenamefont
  {Ni},\ and\ \citenamefont {Xu}}]{Gao2021Nature}%
  \BibitemOpen
  \bibfield  {author} {\bibinfo {author} {\bibfnamefont {A.}~\bibnamefont
  {Gao}}, \bibinfo {author} {\bibfnamefont {Y.-F.}\ \bibnamefont {Liu}},
  \bibinfo {author} {\bibfnamefont {C.}~\bibnamefont {Hu}}, \bibinfo {author}
  {\bibfnamefont {J.-X.}\ \bibnamefont {Qiu}}, \bibinfo {author} {\bibfnamefont
  {C.}~\bibnamefont {Tzschaschel}}, \bibinfo {author} {\bibfnamefont
  {B.}~\bibnamefont {Ghosh}},  \emph {et~al.},\ }\bibfield  {title} {\enquote
  {\bibinfo {title} {{Layer Hall effect in a 2D topological axion
  antiferromagnet}}}, }\href {\doibase 10.1038/s41586-021-03679-w} {\bibfield
  {journal} {\bibinfo  {journal} {Nature}\ }\textbf {\bibinfo {volume} {595}},\
  \bibinfo {pages} {521} (\bibinfo {year} {2021})}\BibitemShut {NoStop}%
\bibitem [{\citenamefont {Chen}\ \emph {et~al.}(2022)\citenamefont {Chen},
  \citenamefont {Sun}, \citenamefont {Gu}, \citenamefont {Hua}, \citenamefont
  {Liu}, \citenamefont {Lu},\ and\ \citenamefont {Xie}}]{ChenR2022NSR}%
  \BibitemOpen
  \bibfield  {author} {\bibinfo {author} {\bibfnamefont {R.}~\bibnamefont
  {Chen}}, \bibinfo {author} {\bibfnamefont {H.-P.}\ \bibnamefont {Sun}},
  \bibinfo {author} {\bibfnamefont {M.}~\bibnamefont {Gu}}, \bibinfo {author}
  {\bibfnamefont {C.-B.}\ \bibnamefont {Hua}}, \bibinfo {author} {\bibfnamefont
  {Q.}~\bibnamefont {Liu}}, \bibinfo {author} {\bibfnamefont {H.-Z.}\
  \bibnamefont {Lu}}, \ and\ \bibinfo {author} {\bibfnamefont {X.~C.}\
  \bibnamefont {Xie}},\ }\bibfield  {title} {\enquote {\bibinfo {title} {{Layer
  Hall effect induced by hidden Berry curvature in antiferromagnetic
  insulators}}}, }\href {\doibase 10.1093/nsr/nwac140} {\bibfield  {journal}
  {\bibinfo  {journal} {Natl. Sci. Rev.}\ }\textbf {\bibinfo {volume} {11}},\
  \bibinfo {pages} {nwac140} (\bibinfo {year} {2022})}\BibitemShut {NoStop}%
\bibitem [{\citenamefont {Jechumt\'al}\ \emph {et~al.}(2026)\citenamefont
  {Jechumt\'al}, \citenamefont {Gueckstock}, \citenamefont {Jasensk\'y},
  \citenamefont {Ka\ifmmode~\check{s}\else \v{s}\fi{}par}, \citenamefont
  {Olejn\'{\i}k}, \citenamefont {Gaerner}, \citenamefont {Reiss}, \citenamefont
  {Moser}, \citenamefont {Kessler}, \citenamefont {De~Luca}, \citenamefont
  {Ganguly}, \citenamefont {Santiso}, \citenamefont {Scheffler}, \citenamefont
  {Z\'azvorka}, \citenamefont {Kuba\ifmmode \check{s}\else
  \v{s}\fi{}\ifmmode~\check{c}\else \v{c}\fi{}\'{\i}k}, \citenamefont
  {Reichlov\'a}, \citenamefont {Schmoranzerov\'a}, \citenamefont
  {N\ifmmode~\check{e}\else \v{e}\fi{}mec}, \citenamefont {Jungwirth},
  \citenamefont {Ku\ifmmode~\check{z}\else \v{z}\fi{}el}, \citenamefont
  {Kampfrath},\ and\ \citenamefont {N\'advorn\'{\i}k}}]{Jechumtal2026PRB}%
  \BibitemOpen
  \bibfield  {author} {\bibinfo {author} {\bibfnamefont {J.}~\bibnamefont
  {Jechumt\'al}}, \bibinfo {author} {\bibfnamefont {O.}~\bibnamefont
  {Gueckstock}}, \bibinfo {author} {\bibfnamefont {K.}~\bibnamefont
  {Jasensk\'y}}, \bibinfo {author} {\bibfnamefont {Z.}~\bibnamefont
  {Ka\ifmmode~\check{s}\else \v{s}\fi{}par}}, \bibinfo {author} {\bibfnamefont
  {K.}~\bibnamefont {Olejn\'{\i}k}}, \bibinfo {author} {\bibfnamefont
  {M.}~\bibnamefont {Gaerner}},  \emph {et~al.},\ }\bibfield  {title} {\enquote
  {\bibinfo {title} {Spin- to charge-current conversion in altermagnetic
  candidate ${\mathrm{ruo}}_{2}$ probed by terahertz emission spectroscopy}},
  }\href {\doibase 10.1103/d17q-1lg4} {\bibfield  {journal} {\bibinfo
  {journal} {Phys. Rev. B}\ }\textbf {\bibinfo {volume} {113}},\ \bibinfo
  {pages} {054439} (\bibinfo {year} {2026})}\BibitemShut {NoStop}%
\bibitem [{\citenamefont {Hayami}\ \emph {et~al.}(2020)\citenamefont {Hayami},
  \citenamefont {Yanagi},\ and\ \citenamefont {Kusunose}}]{Hayami2020PRB}%
  \BibitemOpen
  \bibfield  {author} {\bibinfo {author} {\bibfnamefont {S.}~\bibnamefont
  {Hayami}}, \bibinfo {author} {\bibfnamefont {Y.}~\bibnamefont {Yanagi}}, \
  and\ \bibinfo {author} {\bibfnamefont {H.}~\bibnamefont {Kusunose}},\
  }\bibfield  {title} {\enquote {\bibinfo {title} {Bottom-up design of
  spin-split and reshaped electronic band structures in antiferromagnets
  without spin-orbit coupling: Procedure on the basis of augmented
  multipoles}}, }\href {\doibase 10.1103/PhysRevB.102.144441} {\bibfield
  {journal} {\bibinfo  {journal} {Phys. Rev. B}\ }\textbf {\bibinfo {volume}
  {102}},\ \bibinfo {pages} {144441} (\bibinfo {year} {2020})}\BibitemShut
  {NoStop}%
\end{thebibliography}%

\end{document}